\documentclass[superscriptaddress,twocolumn,amsmath,amssymb,aps,pre,floatfix]{revtex4}

\usepackage[english]{babel}
\usepackage[utf8]{inputenc}
\usepackage[T1]{fontenc}
\usepackage{booktabs}
\usepackage{xr}
\usepackage{float}
\usepackage{natbib}
\usepackage{units}
\usepackage{graphicx}
\usepackage{dcolumn}
\usepackage{bm}
\usepackage{hyperref}
\hypersetup{colorlinks=true, linktoc=all, linkcolor=blue, linktocpage, citecolor=blue}

\begin{document}

\title{Curvature-driven growth and interfacial noise in the voter model with self-induced zealots}

\author{Luís Carlos F. Latoski}
\email{luis.latoski@ufrgs.br}
\affiliation{Instituto de Física, Universidade Federal do Rio Grande do Sul, CEP 91501-970, Porto Alegre - RS, Brazil} 

\author{W. G. Dantas}
\email{wgdantas@id.uff.br}
\affiliation{Departamento de Ciências Exatas, EEIMVR, Universidade Federal Fluminense, CEP 27255-125, Volta Redonda - RJ, Brazil}

\author{Jeferson J. Arenzon}
\email{arenzon@if.ufrgs.br}
\affiliation{Instituto de Física, Universidade Federal do Rio Grande do Sul, CEP 91501-970, Porto Alegre - RS, Brazil} 
\affiliation{Instituto Nacional de Ciência e Tecnologia - Sistemas Complexos, Rio de Janeiro RJ, Brazil}

\date{\today}

\begin{abstract}
We introduce a variant of the voter model in which agents may have different degrees of confidence on their opinions. Those with low confidence are normal voters whose state can change upon a single contact with a different neighboring opinion. However, confidence increases with opinion reinforcement and, above a certain threshold, these agents become zealots that do not change opinion. We show that both strategies, normal voters and zealots, may coexist, leading to a competition between two different kinetic mechanisms: curvature-driven growth and interfacial noise. The kinetically constrained zealots are formed well inside the clusters, away from the different opinions at the surfaces that help keep the confidence not so high. Normal voters concentrate in a region around the interfaces and their number, that is related with the distance between the surface and the zealotry bulk, depends on the rate the confidence changes. Despite this interface being rough and fragmented, typical of the voter model, the presence of zealots in the bulk of these domains, induces a curvature-driven dynamics, similar to the low temperature coarsening behavior of the non-conserved Ising model after a temperature quench. 
\end{abstract} 

\maketitle

\maketitle

\section{Introduction}

Consensus, in physical models of opinion dynamics~\cite{CaFoLo09,Baronchelli18,Redner19},  may be achieved locally, within a given subgroup, or globally, with all interacting agents agreeing on a common position.
Understanding the process of formation and the probability of attaining a consensus, how to enforce it and why it is sometimes prevented is important to uncover its underlying, universal mechanisms. 
An example is the convergence of results in some scientific fields (vaccines, climate, etc) that, despite being widely accepted in the scientific community~\cite{LyHoPe21,StBrJa21}, do not always lead to evidence-based public policies.
Indeed, in actual situations, the agreement with other individuals may depend on several factors (e.g., the local network, intensity of noise, new evidences, propaganda, fake-news, self-confidence and other psychological reinforcement mechanisms, etc). 
However, in simple systems like the binary Voter model (VM) and related models of language competition, the process is simplified and considered as an ordering one, in which each agent aligns its opinion with one of its neighbours. 
In an infinite, regular lattice of dimension $d\leq 2$, the consensus in the VM is an absorbing state (bulk noise is absent) and always attained albeit with very different time dependencies~\cite{Krapivsky92,Liggett99,FrKr96,BeFrKr96}.
In 2d, the geometry we consider here, the growth of order by coarsening, in the absence of surface tension, is not driven by the curvature of the interfaces but by its noise~\cite{DoChChHi01,TaCuPi15}. 

In the original VM, agents have no confidence whatsoever and opinions may change upon a single contact with a different position.
The possibility of having a strong confidence is among the many modifications introduced to better describe more complex social phenomena~\cite{CaFoLo09,Baronchelli18,Redner19}. 
These confident agents, or zealots, may never change opinion, or change it in spite of the opinions of their neighbours~\cite{Mobilia03,MoGe05,MoPeRe07,GaJa07,CoCa16}, introducing some disorder in the system that, besides interfering on how the local consensus groups grow, may even prevent the system from attaining a global consensus. 
When intermediate levels of confidence are allowed, the zealot state may be transient~\cite{DaCa07,StTeSc08,StTeSc08b,DaGa08,LaSaBl09,VoRe12,MaGa13,WaLiWaZhWa14,BrTo15,VeVa18} and opinions are kept while the necessary number of contacts is not attained (complex contagion).
This is akin to an annealed disorder and represents the inertia in the process of changing opinion, associated with a reinforcement mechanism that makes positions stronger or weaker. 
The necessity of multiple contacts prior to a change of state is similar to sampling the local field by interacting with a larger number of closest neighbors~\cite{DaCa07}.
This noise reduction mechanism induces an effective surface tension~\cite{CaEgSa06,DaCa07,DaGa08,CaBaLo09,VoRe12,ZhLiKoSz14,DoSrSzKo16,RoSe17,VeVa18,MuBiPa20} and some properties become analogous to those of the low temperature coarsening of the Ising model with non-conserved order parameter, in the Allen-Cahn (Model A) universality class~\cite{Bray94}.

We propose an alternative model in which, instead of zealotry being an inherited characteristic of all agents, it may develop depending on the individual previous history. 
While its confidence is low, the agent behaves as a normal voter.
However, above a given threshold its opinion freezes and it becomes a kinetically constrained zealot.
This self-induced  disorder may be either irreversible (quenched) or, when the confidence keeps evolving,  reversible (annealed).  
A reversible, or transient zealot, needs multiple interactions with the opposite opinion to reset its confidence and once again be able to change its opinion, a complex contagion process, differently from the simple, single contact process for normal agents.
During the dynamics, clusters of agents with a common opinion form, grow and compete towards the consensus state.
The reinforcement process between agents with the same opinion leads, after multiple interactions, to the formation of zealots in the bulk of these clusters.
Because of the constant flipping that occurs close to the surface where both opinions coexist, the confidences are repeatedly reset,  the agents tend to be normal voters and, consequently, the surface is very rough and fragmented.
Below the actual interface there is another one, internal, separating the bulk zealots from normal voters, all with the same opinion.
Close to this secondary surface, normal voters that are close to zealots have a persistent neighborhood that induce an increase in their confidence, eventually increasing the probability of becoming zealots themselves.
This seems to be the mechanism responsible for the effective surface tension of the internal surface, that behaves as a frame structure for the external one.
An important question is whether the internal surface is enough to turn the dynamics into a curvature-driven one, in spite of the interface still being rough as in the original VM.
Moreover, what are the consequences for the probability of attaining a consensus? How does the approach to the stationary state changes, depending on the parameters of the model? How does the geometric properties of the opinion clusters (neighboring agents with the same opinion) differ from those of the pure VM?
These are some of the questions that we try to answer in the following sections.


\section{The Model}
\label{sec.model}

The state of each agent is characterized by two variables, $(\sigma_i,\eta_i)$.
The binary opinion is represented by the discrete variable $\sigma_i=\pm 1$, where $i=1,\ldots,N$. 
The total number of agents, $N$, corresponds to the sites either in a 1d ring or in a 2d square lattice where $N=L^2$. 
Each opinion is associated with some individual degree of confidence, which is described by the continuous variable $\eta_i\geq0$.
It depends on the previous history  of contacts and evolves after each interaction. 
When $\eta_i$ attains the threshold $\phi$ (which is set to $\phi=1$), the agent becomes refractory to the opinions of its neighbors and $\sigma_i$ is temporarily frozen, a form of self-induced disorder.
However, $\eta_i$ keeps evolving, and when it get smaller than $\phi$, that agent becomes once again susceptible to the opposite opinions of its nearest neighbors and $\sigma_i$ may change. We will refer to transient zealots simply by zealots, while the other agents will be said normal.

In a Monte Carlo step (MCS), $N$ attempts of updating randomly chosen agents are performed.
Two agents are selected, $i$ and one of its nearest neighbors $j$, whose states are, respectively,  $(\sigma_i,\eta_i)$ and $(\sigma_j,\eta_j)$ at time $t$. 
If their opinions differ, $\sigma_i\neq\sigma_j$, and $\eta_i<\phi$, the non-zealot focal site changes its opinion and aligns with $j$:
\begin{equation}
    \label{eq.opinionchange}
    \sigma_i \longrightarrow \sigma_j, \text{  if  } \eta_i<\phi.
\end{equation}
Although zealots, obviously, do not change their opinions, the confidence of both $i$ and $j$ are updated, in this case, accordingly with
\begin{align}
    &\label{eq2}\eta_i   \longrightarrow \eta_i/\gamma \\
    &\label{eq3}\eta_j   \longrightarrow \eta_j + \Delta \eta,
\end{align}
where $\gamma$ and $\Delta\eta$ are positive parameters.
The fact of $i$ being confronted with a different opinion is enough to change its confidence by the rescaling factor $\gamma$. 
For intermediate values, $1<\gamma<\infty$, $\eta_i$ continuously decreases and zealots eventually may become normal once again. 
When $\gamma\leq 1$, $\eta_i$ does not decrease and becoming a zealot is an irreversible process that may prevent the system from attaining the consensus.
This mimics the reinforcement observed in conspiracy theories and among negationists.
The confidence of the neighbour $j$, on the other hand, always increases by $\Delta\eta$ because it had the opportunity to expose its opinion to a neighbour. 
Finally, when both agents have the same opinion, $\sigma_i=\sigma_j$, the mutual reinforcement is positive and both confidences increase:
\begin{equation}
\eta_{i,j}   \longrightarrow \eta_{i,j} + \Delta \eta .
\label{eq6}
\end{equation}

We study the above competing mechanisms in the extreme cases $\gamma=1$ and $\gamma\to\infty$. 
If $\gamma\to\infty$ and $\sigma_i\neq\sigma_j$, the confidence of the focal agent is always reset, i.e., $\eta_i$ instantly becomes 0. Thus, whatever the degree of zealotry, only two steps may be enough to any agent change its opinion: in the first interaction the confidence is reset and in the next step the opinion may be updated. In this way, the model combines simple and complex contagion where  single or multiple exposures are necessary, respectively, to induce a change of opinion. 
The other case, $\gamma=1$ corresponds to the irreversible limit where the confidence never decreases and the system gets frozen once all $\eta_i$ become larger than $\phi$. 
Irrespective of the value of $\gamma$, in the initial steps of the dynamics, when zealots have not yet been created, the model is equivalent to the standard VM, but there is a $\gamma$-dependent timescale when it crosses over to a new behavior. 
The main objective of this paper is, indeed, to describe and understand how the behavior is affected by the presence of irreducible agents.
    
\section{Results}

\subsection{1d}
\label{sec.results1d}

Initially, the variables $\{\sigma_i,\eta_i\}$ characterizing the state of all $N$ sites are randomly assigned: opinions are chosen with equal probability of being $\sigma_i=\pm 1$ while all agents share the same level of confidence, $\eta_i=0$. Although we have performed simulations with several different sizes of the 1d ring, we only present  the results for $N=10^5$, which is enough to reduce finite size effects. All results are averaged over 100 samples.
Besides measuring the density of neighboring sites whose opinions differ, $\rho(t)$ (the fraction of defects), we also consider the persistence $P(t)$, the fraction of sites that have not changed their initial state up to time $t$~\cite{BrMaSc13}.

\begin{figure}[htb]
\includegraphics[width=\columnwidth]{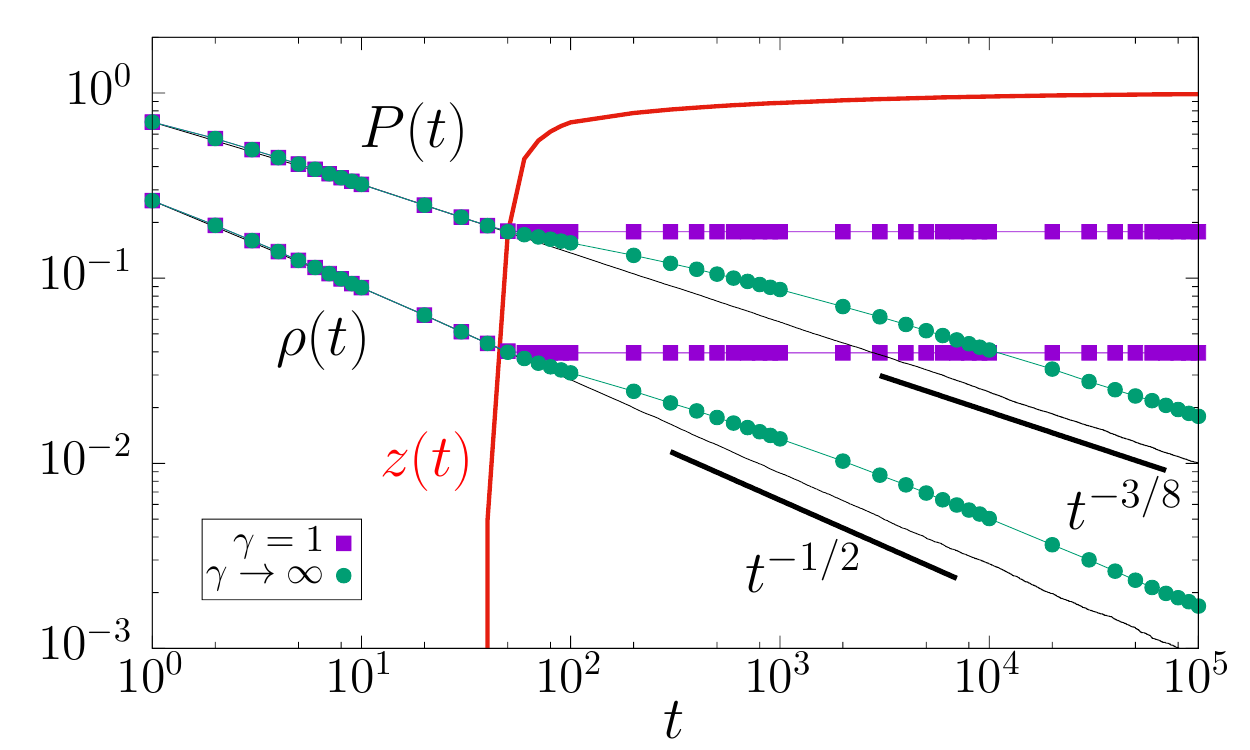}
\caption{Persistence $P(t)$ and density of defects $\rho(t)$ in 1d for $\Delta \eta = 10^{-2}$, $\gamma=1$ (purple symbols) and $\gamma\to\infty$ (green symbols). The solid black curves correspond to the pure 1d VM where $P(t)\sim t^{-3/8}$~\cite{Derrida95,DeHaPa95} and $\rho(t) \sim t^{-1/2}$. The red curve shows the very fast increase of the fraction of zealots, $z(t)$. There are three different regimes. In the initial one, the system follows the VM behavior. It is followed by an intermediate regime that starts when zealots first appear and the curves deviate from the VM behavior. Eventually, in the third regime, the system resumes the exponents characterizing the VM behavior.}
\label{fig.1d-1}
\end{figure}

Fig.~\ref{fig.1d-1} shows the temporal evolution of both $P(t)$ and $\rho(t)$. 
Since zealots only appear later in the dynamics (continuous red line), the initial trend is the same as the VM: $P(t)\sim t^{-3/8}$~\cite{Derrida95,DeHaPa95} and $\rho(t)\sim t^{-1/2}$~\cite{FrKr96}. 
However, along with the fast increase in the density of zealots, we observe deviations from the VM behavior. 
For $\gamma=1$, once zealots are created, the system ends in a frozen configuration with two or more compact blocks of opposite opinions, consensus is never achieved.
On the other hand, for $\gamma\to\infty$, both $P(t)$ and $\rho(t)$ slow down their decrease during a transient interval, soon resuming the VM exponent at longer times, albeit with a larger coefficient. 

\begin{figure}[htb]
\includegraphics[height=.3\columnwidth,width=.3\columnwidth,trim={.025cm .025cm .025cm .025cm},clip]{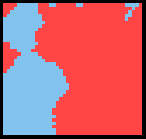}
\includegraphics[height=.3\columnwidth,width=.3\columnwidth,trim={.025cm .025cm .025cm .025cm},clip]{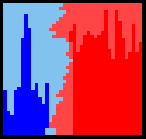}
\includegraphics[height=.3\columnwidth,width=.3\columnwidth,trim={.025cm .025cm .025cm .025cm},clip]{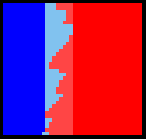}
\caption{Snapshots for the 1d case with $\gamma\to\infty$ and $\Delta \eta = 10^{-2}$, showing the temporal evolution in the three different regimes (time goes from top to bottom and only part of the lattice is shown). Dark colors are used for zealots while normal voters ($\eta_i<\phi$) have light colors. In the initial regime (left panel), the dynamics is indistinguishable from the original VM and several small domains coexist. The middle panel shows the intermediate regime where many of these domains already coalesced and the first zealots appear away from the interfaces. Finally, the right panel illustrates the long time behavior where there are two compact blocks of zealots with opposite opinion and, in the middle, a region with the normal agents.}
\label{fig.1d-2}
\end{figure}

The different temporal regimes are illustrated in the snapshots of Fig.~\ref{fig.1d-2}.
In the initial regime, left panel, zealots are absent and the dynamics is the same as the VM.
The system is divided in small domains that coalesce once two walls collide~\cite{Hinrichsen00}.
In the intermediate regime, middle panel, zealots appear (dark colors) in the interior of domains.
Normal agents of both opinions  (light colors) become confined between compact blocks of zealots, characterizing the late stage of the dynamics, right panel. 
Inside these stripes, the normal agents follow the VM and once the moving border (where light red and light blue are neighbors) gets closer to zealots, decreasing their confidence, the width of the stripe may change.  
Because of this further step necessary to unblock the zealots, the spreading is slower than in the VM.
Small domains last longer and both the persistence and the number of interfaces are relatively larger.

\begin{figure}[ht]
\includegraphics[width=\columnwidth]{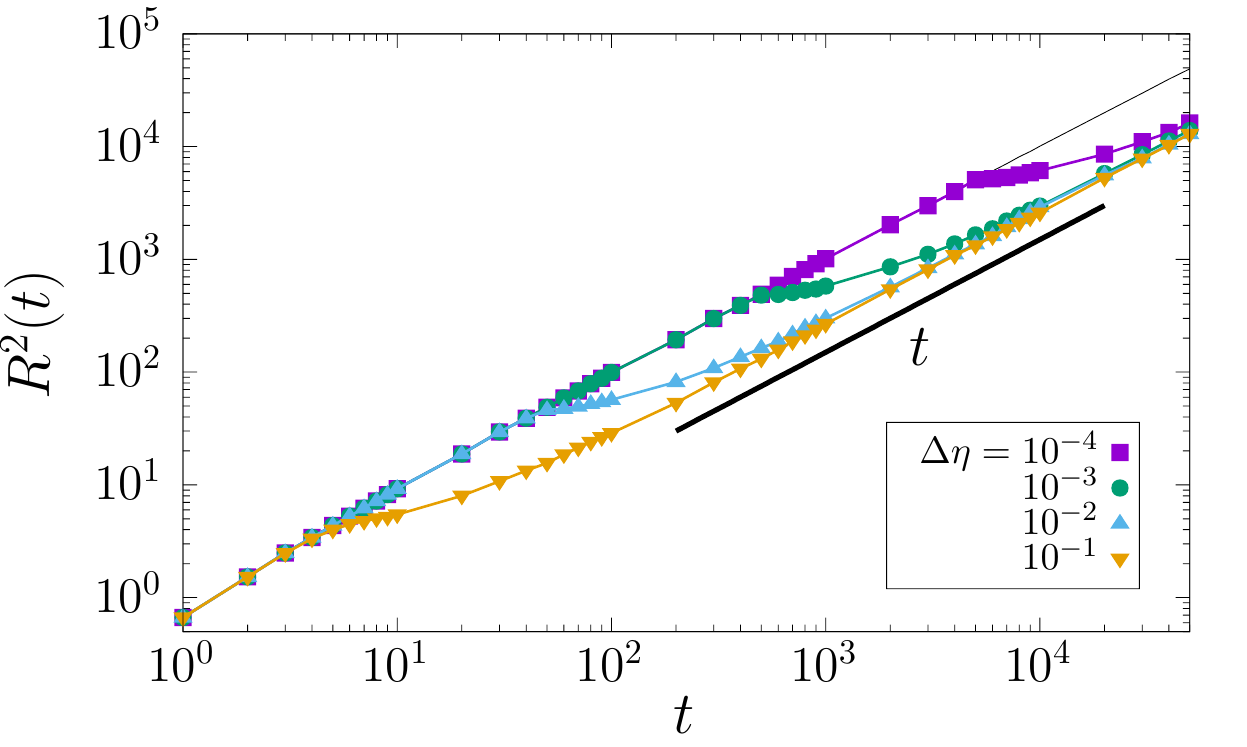}
\caption{Mean square displacement $R^2(t)$ of the single wall, for different values of $\Delta\eta$. The thin black line is the linear, diffusive behavior of the VM. Both at short and long times, the behavior is diffusive, $R^2(t)\sim t$, for all values of $\Delta\eta$. There is, however, an intermediate, sub-diffusive regime in which the curves depart from and, later, return to the linear behavior. 
}
\label{fig.1d-msd}
\end{figure}

The coalescence of domains, in 1d, is driven by the diffusive behavior of the domain walls~\cite{Hinrichsen00}. 
In order to better understand how the presence of zealots affects such behavior, we consider an initial state in which there is a single domain wall, located at $x(0)$, dividing the system in two blocks, one with each $\sigma_i=\pm 1$ state. The boundary conditions are open, $\gamma\to\infty$ and all sites start with $\eta_i=0$. 
The mean square displacement $R^2(t)=[x(t)-x(0)]^2$, where $x(t)$ is the location of the interface at time $t$, is shown in Fig.~\ref{fig.1d-msd} for different values of $\Delta\eta$.  
In all cases, two timescales are present. 
The initial behavior is purely diffusive, as in the VM (thin black line), and  $R^2(t)\sim t$.
The smaller $\Delta\eta$ is, the longer it will take for the system to deviate from the original VM behavior, becoming sub-diffusive.
This deviation occurs at an intermediate time that behaves as $(\Delta\eta)^{-1}$, when zealots first appear.
On a longer timescale, that also goes as $(\Delta\eta)^{-1}$, the diffusive behavior is resumed. 
At the late stage of the dynamics, all activity is confined to the stripe between the two blocks of zealots and the interface evolution depends on the unblocking of the neighboring zealots, what occurs on a longer timescale.
The overall behavior is reminiscent of glassy systems, with a fast timescale associated with the Brownian motion inside the cage formed by neighboring particles and a larger timescale related with the slow restructuring of the cages themselves.

\subsection{2d}
\label{sec.results2d}

In the extreme case $\gamma=1$, as mentioned above, the confidence $\eta_i$ never decreases and the creation of zealots is irreversible, this self-induced disorder is thus quenched.
If $\Delta\eta$ is large enough, the compact domains of zealots grow very fast until colliding with the neighboring domains.
Thus, as in 1d, the bi-dimensional system with $\gamma=1$ attains a frozen state without normal agents and consensus is avoided.
We focus here, instead, on the $\gamma\to\infty$ case, where the confidence if fully reset after a single contact with a different opinion.
For intermediate values of $\gamma$ the system behavior interpolates between these two extremes. The results discussed below were averaged over, at least, 1000 samples.

\begin{figure}[htb]
    \includegraphics[width=.3\columnwidth]{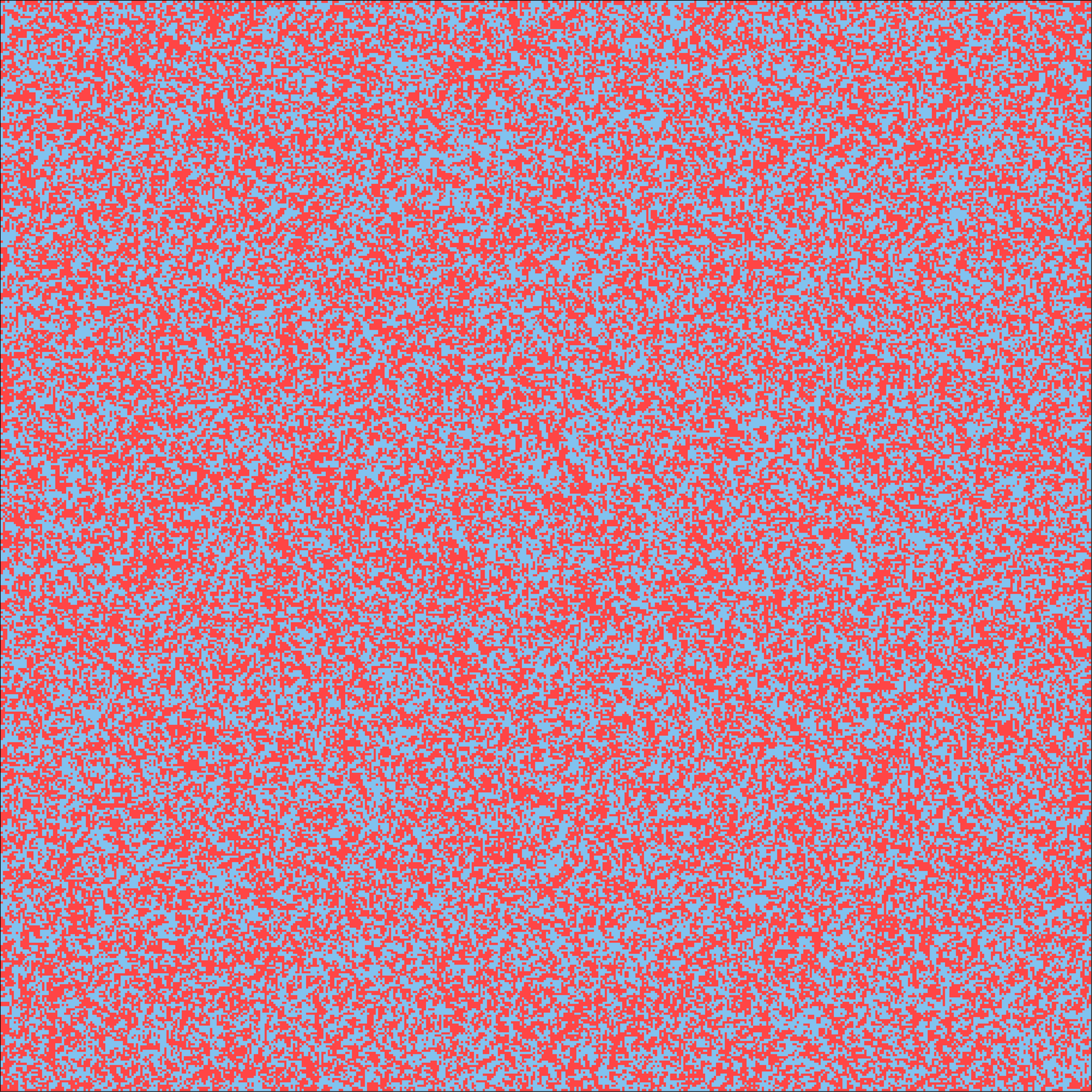}
    \includegraphics[width=.3\columnwidth]{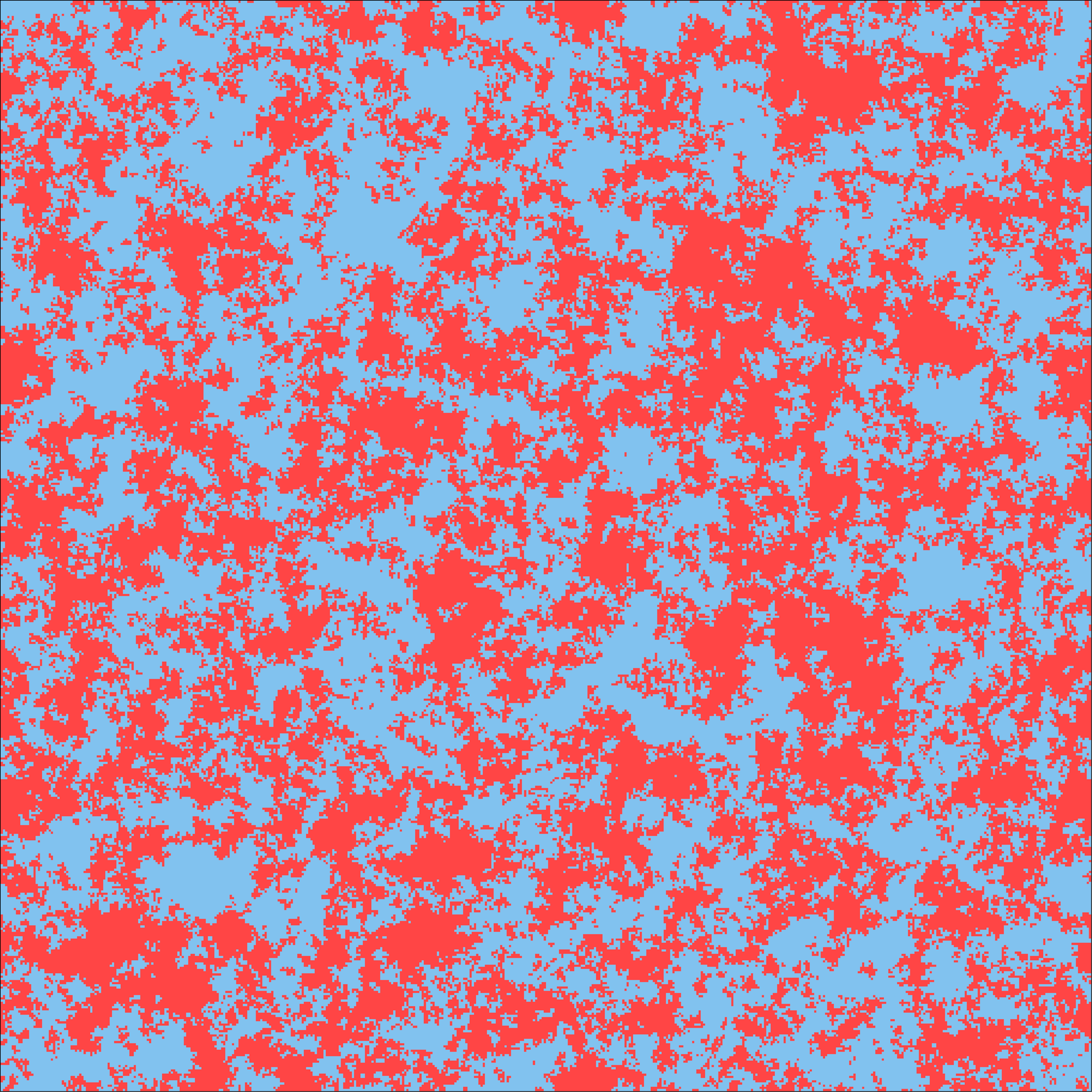}
    \includegraphics[width=.3\columnwidth]{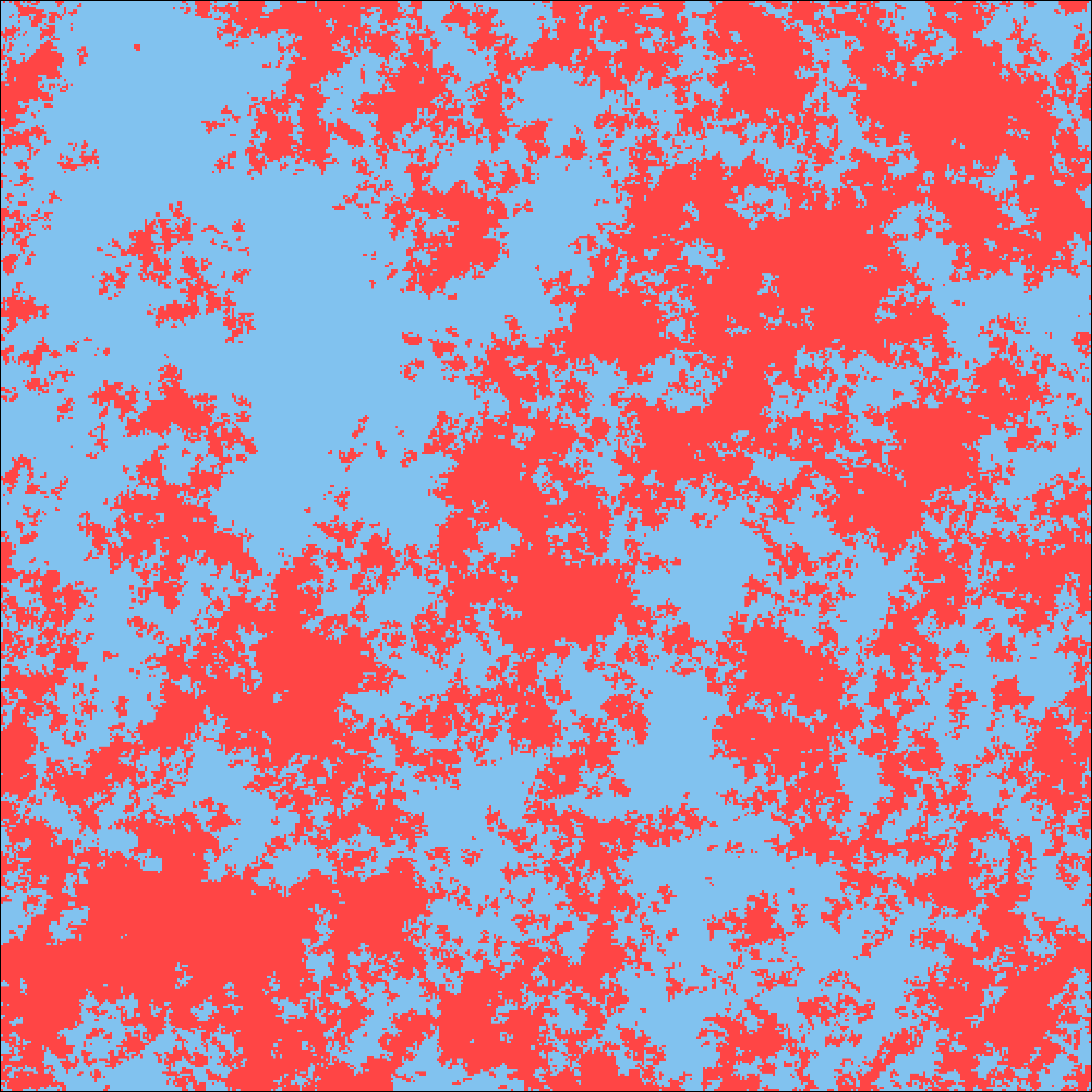}\\
    \vspace{.1cm}
    \includegraphics[width=.3\columnwidth]{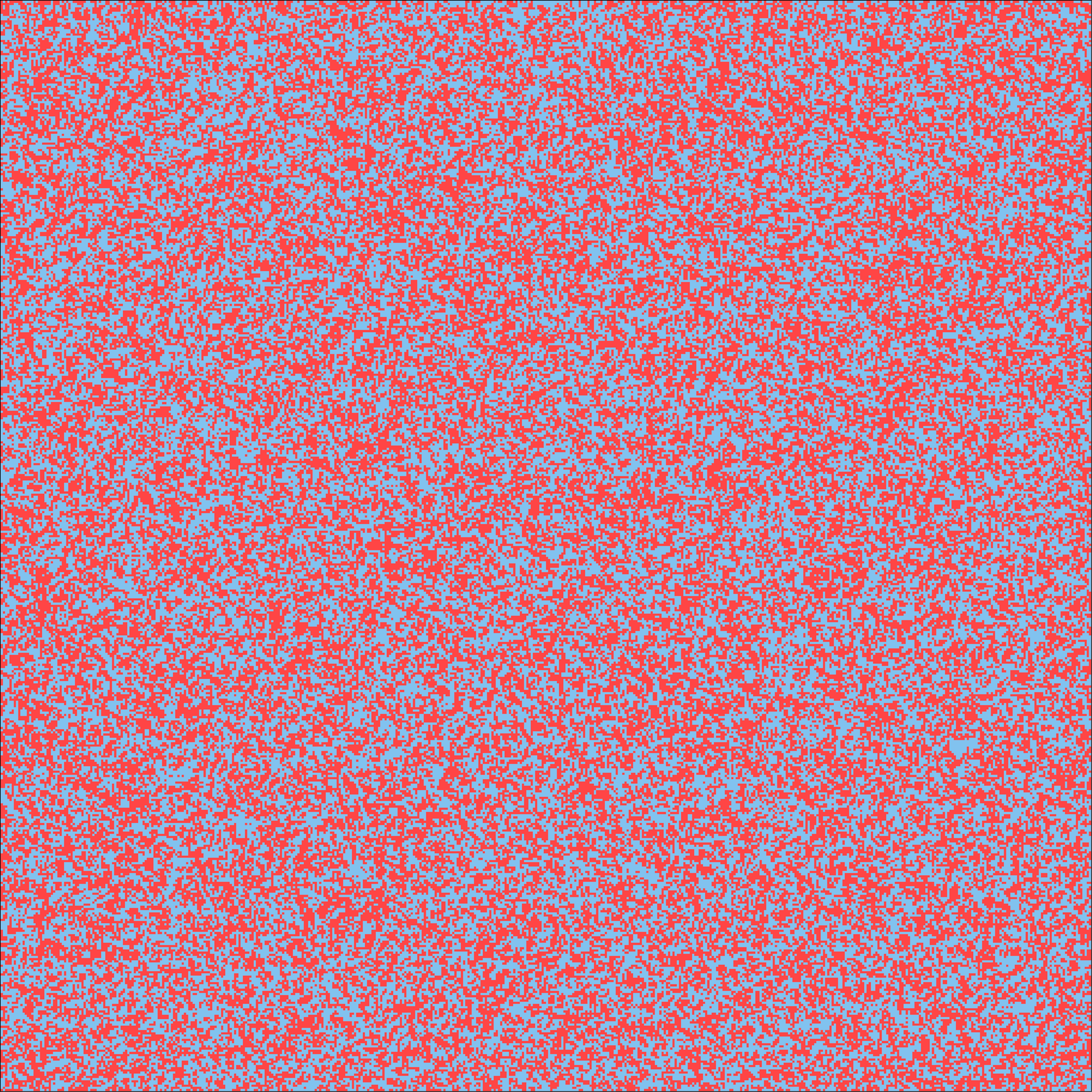}
    \includegraphics[width=.3\columnwidth]{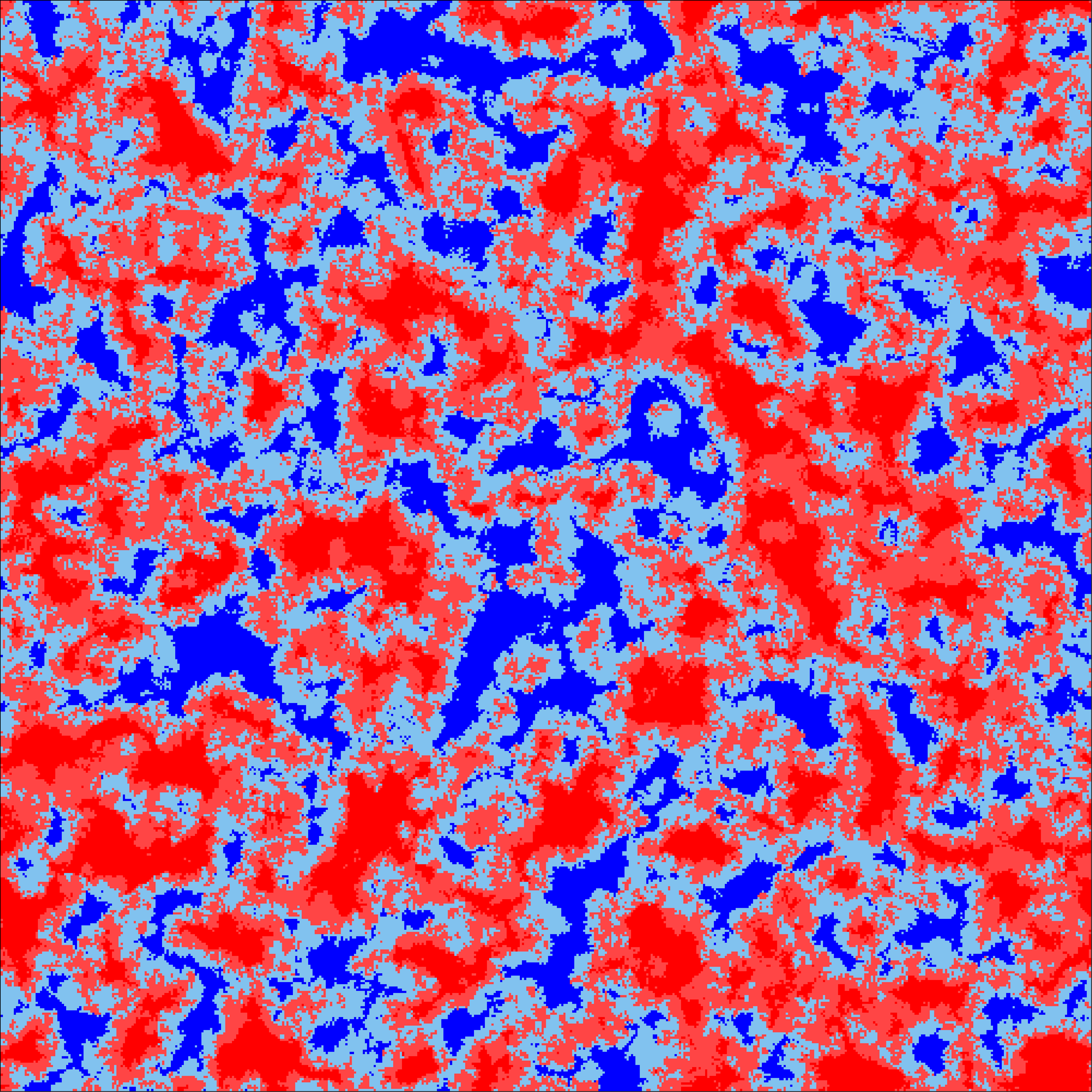} 
    \includegraphics[width=.3\columnwidth]{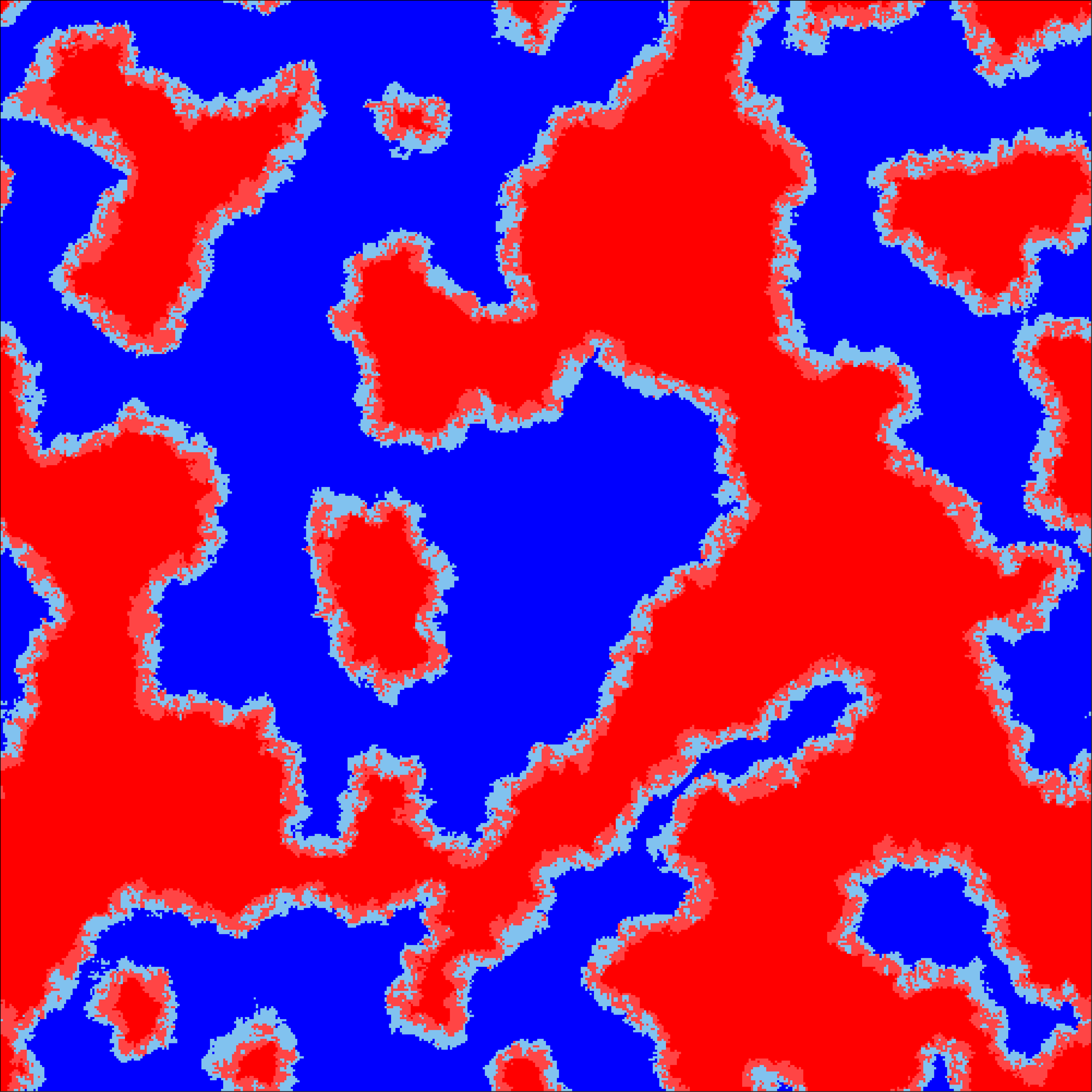}\\
    \vspace{.1cm}
    \includegraphics[width=.3\columnwidth]{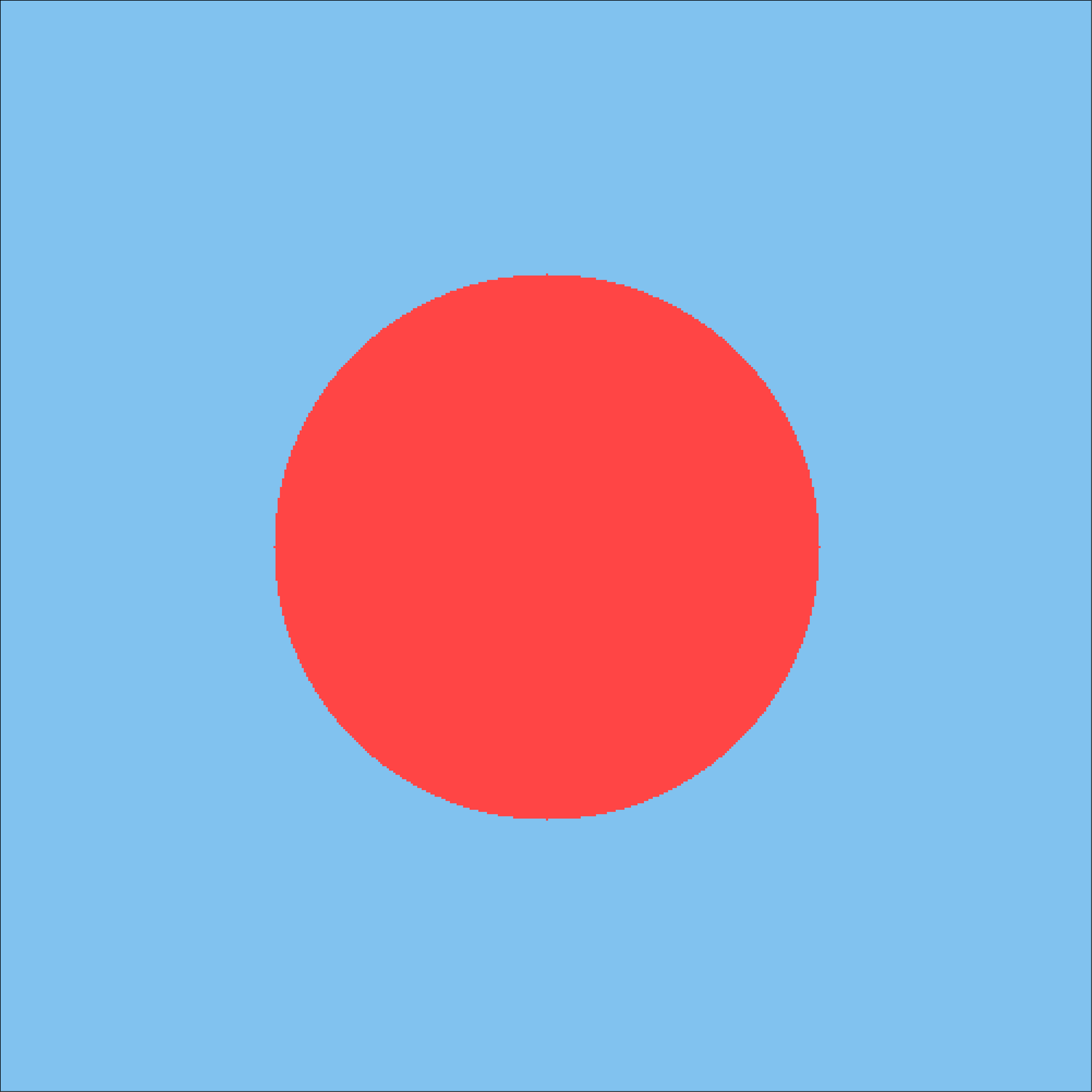}
    \includegraphics[width=.3\columnwidth]{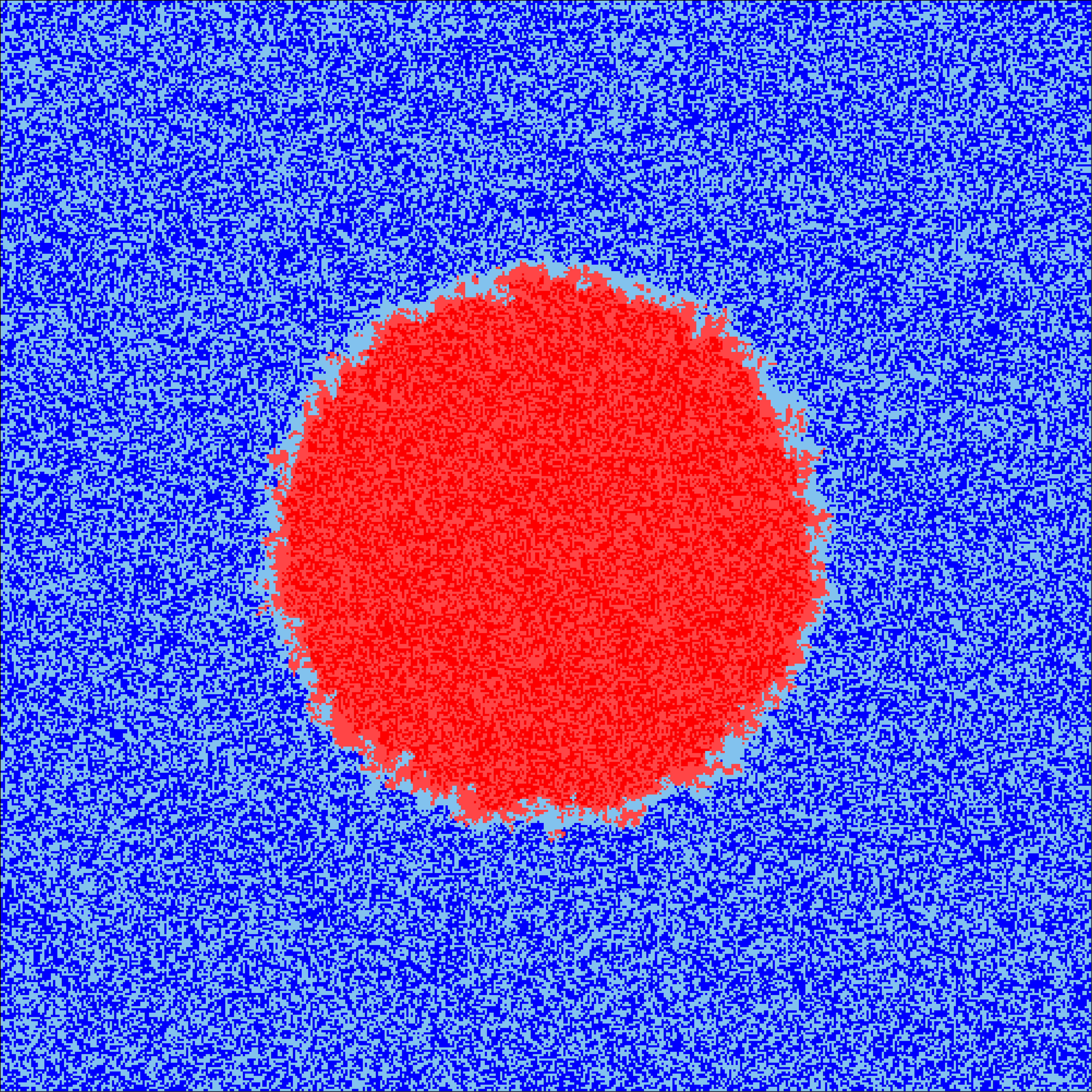} 
    \includegraphics[width=.3\columnwidth]{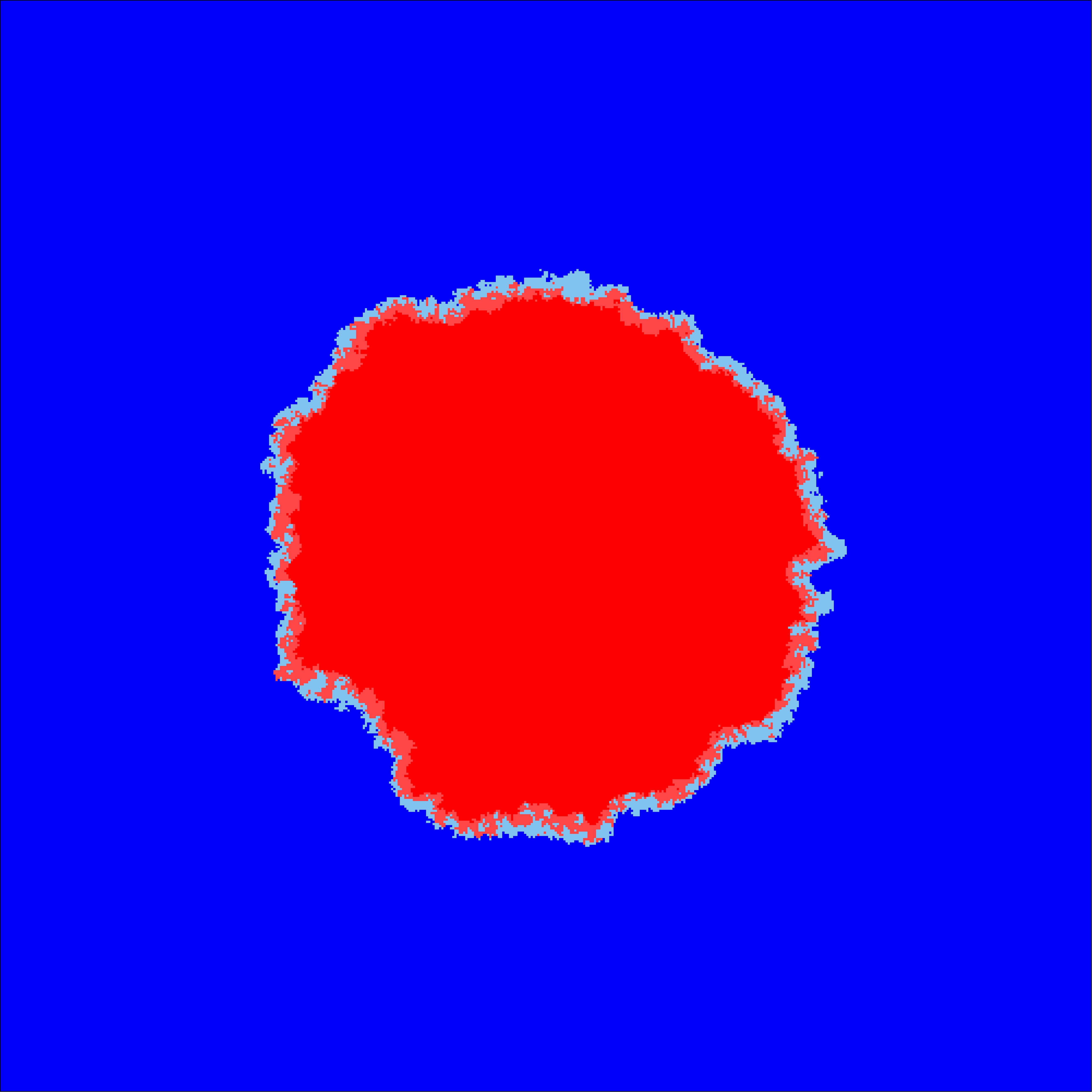}
   \caption{Snapshots at different times for the 2d VM (top row) and our model (middle and bottom row) with $\gamma\to\infty$ and $\Delta\eta=10^{-2}$. Each color is a different opinion, the darker versions indicating zealots while light colors are for normal agents. The onset of zealots induce an effective surface tension and the bulk dynamics becomes curvature-driven. Thus, while the interfaces between normal agents with opposite opinions (light red and light blue) are still rough as in the original VM model, the internal walls, between normal and zealots agents with the same opinion, are smoother. Instead of a random initial state, in the bottom row we consider all agents with one opinion inside a circle, surrounded by the other opinion.
   }
   \label{fig.snap01}
\end{figure}

\begin{figure}[ht!]
    \includegraphics[height=.3\columnwidth,width=.3\columnwidth,trim={21.5cm 21.5cm 21.5cm 21.5cm},clip]{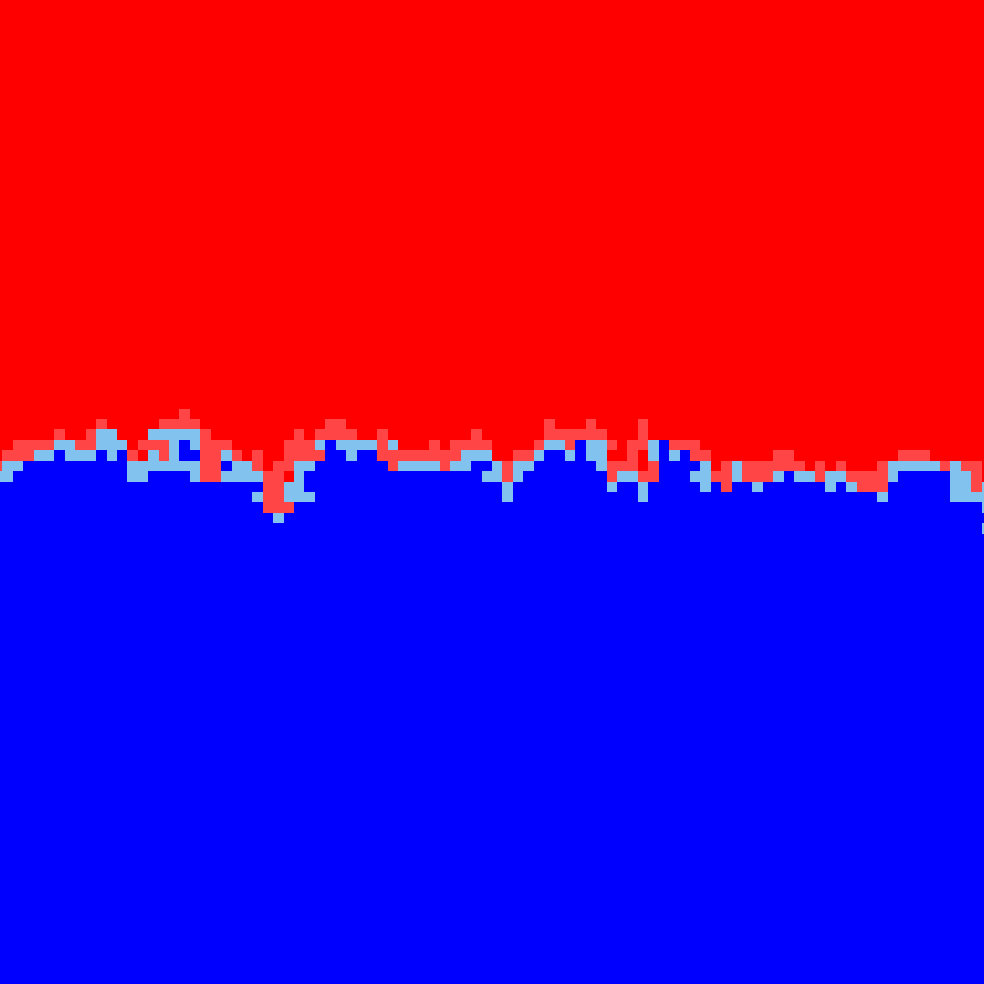}
    \includegraphics[height=.3\columnwidth,width=.3\columnwidth,trim={21.5cm 21.5cm 21.5cm 21.5cm},clip]{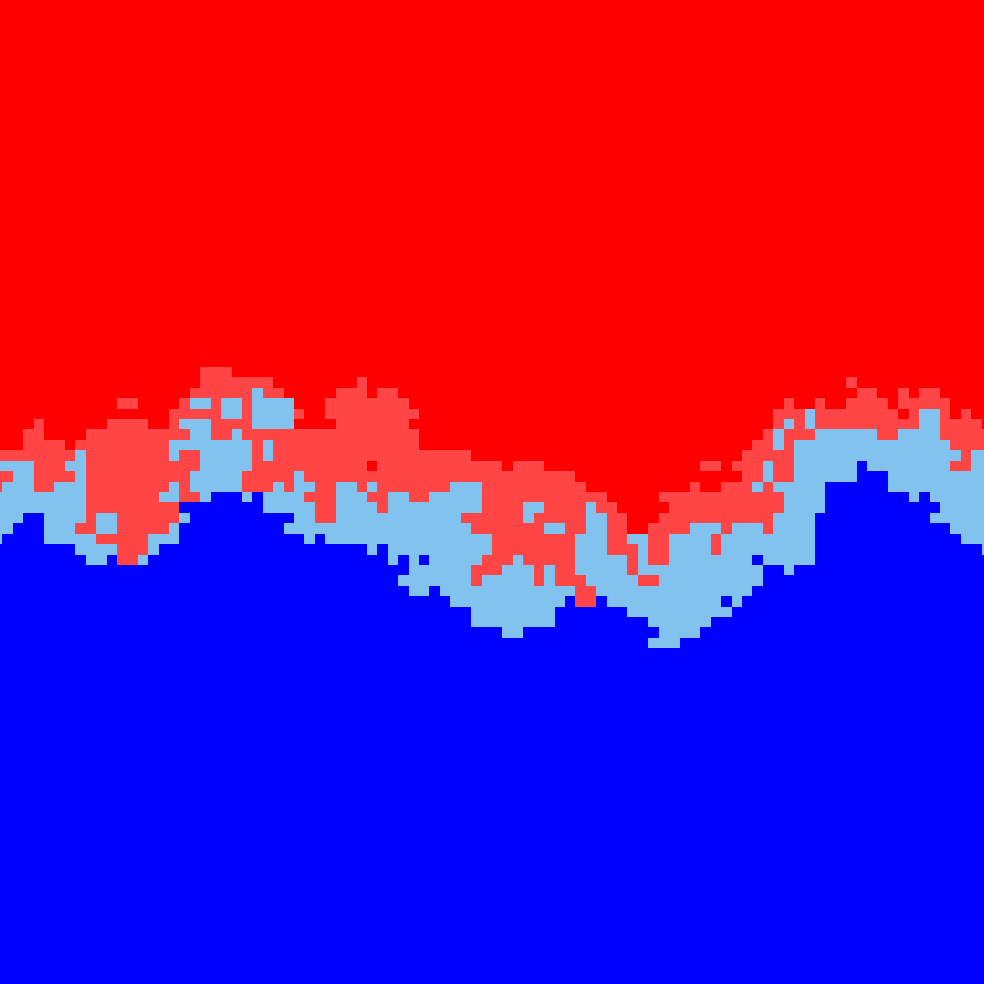}
    \includegraphics[height=.3\columnwidth,width=.3\columnwidth,trim={21.5cm 21.5cm 21.5cm 21.5cm},clip]{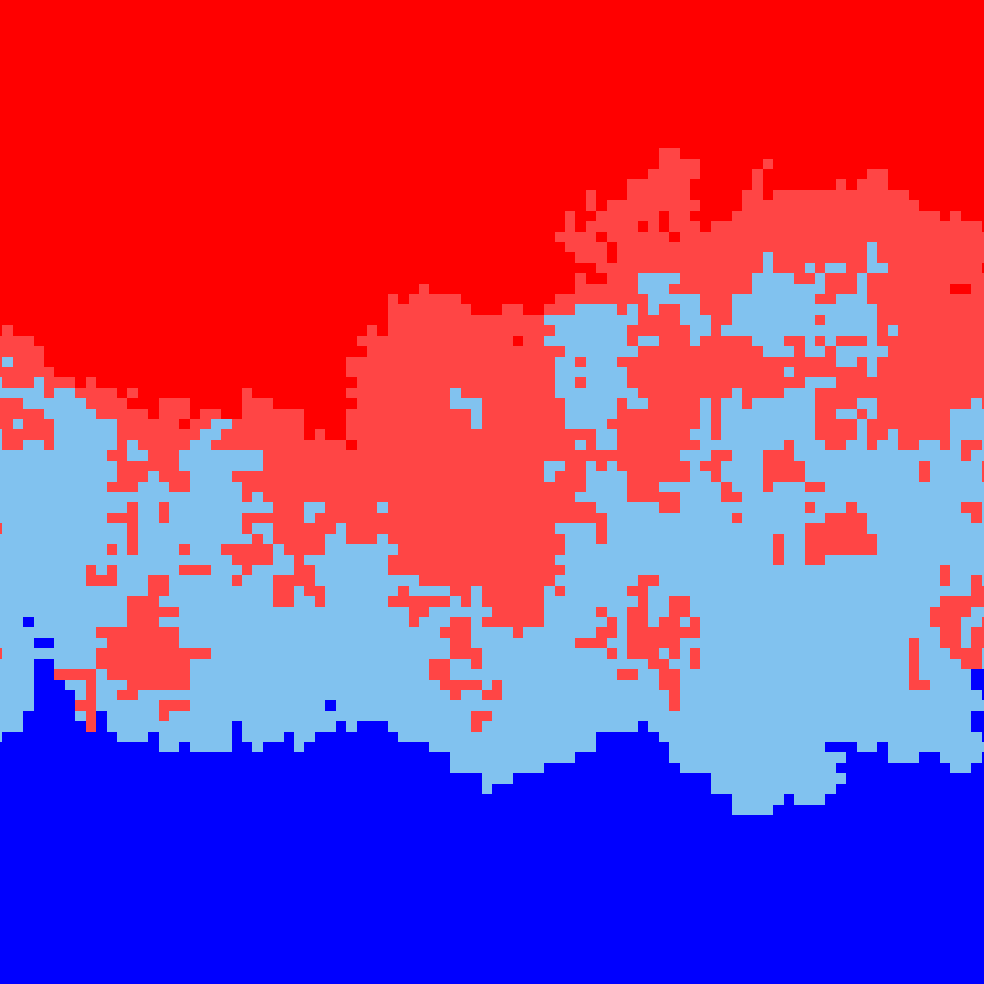}
    \caption{Snapshots showing how the VM stripe develops, after \unit[1000]{MCS}, from a specially prepared initial state with two equal regions of opposite opinions in the $\gamma\to\infty$ case. Only the central part of the system is shown. From left to right the values of $\Delta\eta$ are, respectively: 1, $10^{-1}$ and $10^{-2}$. Notice that the active region gets wider as $\Delta\eta$ decreases with the presence of small domains, without zealots, fully embedded in the other opinion.}
    \label{fig.wall}
\end{figure}

Starting from a random initial state (left panel), the behavior for $\gamma\to\infty$ is illustrated in Fig.~\ref{fig.snap01}, middle row. 
For comparison, the evolution of the original VM is shown in the top row. 
Deep inside the domains, certainty builds up and the agents become zealots, creating an internal interface between bulk zealots and normal agents.
The presence of zealots induces an effective surface tension and this interior interface gets smoother. The dynamics becomes curvature-driven, similar to those of the out-of-equilibrium 2d Ising model as shown in the snapshots in the middle and low rows of Fig.~\ref{fig.snap01}.
As will be shown below, the analogy with the model A dynamical universality class~\cite{Bray94}, to which the out-of-equilibrium Ising model belongs, goes beyond these visual similarities.
Being curvature-driven, the circular domain shrinks~\cite{KaWeDe91,KaLo05,CeLo07,ArCuPi15}, with a much reduced fragmentation when compared with a similar condition for the VM~\cite{DoChChHi01}.
Whatever the initial condition, the external interfaces remain rough at all times because normal agents of both opinions get confined in the VM superficial stripe.
The fluctuations of the main interface cause a constant flipping that keeps the certainties below the threshold in this region, setting the average distance between the external and internal interfaces.
The width of the VM stripe depends on both $\Delta\eta$ and $\gamma$, as illustrated in Fig.~\ref{fig.wall} for an initial state with a flat interface within two equal sized domains. 
For large values of $\Delta\eta$ zealots form very fast and the dynamics gets blocked close to the initial state (notice that the initial state is absorbing for the IM0), left panel. 
As $\Delta\eta$ decreases, the VM stripe becomes wider, the roughness of the interface increases as well as the number of fragmented clusters (middle and right panels). 
Of course, when $\Delta\eta\to 0$, no zealot is formed and the VM stripe is the whole system.

\begin{figure}[ht]
    \includegraphics[width=\columnwidth]{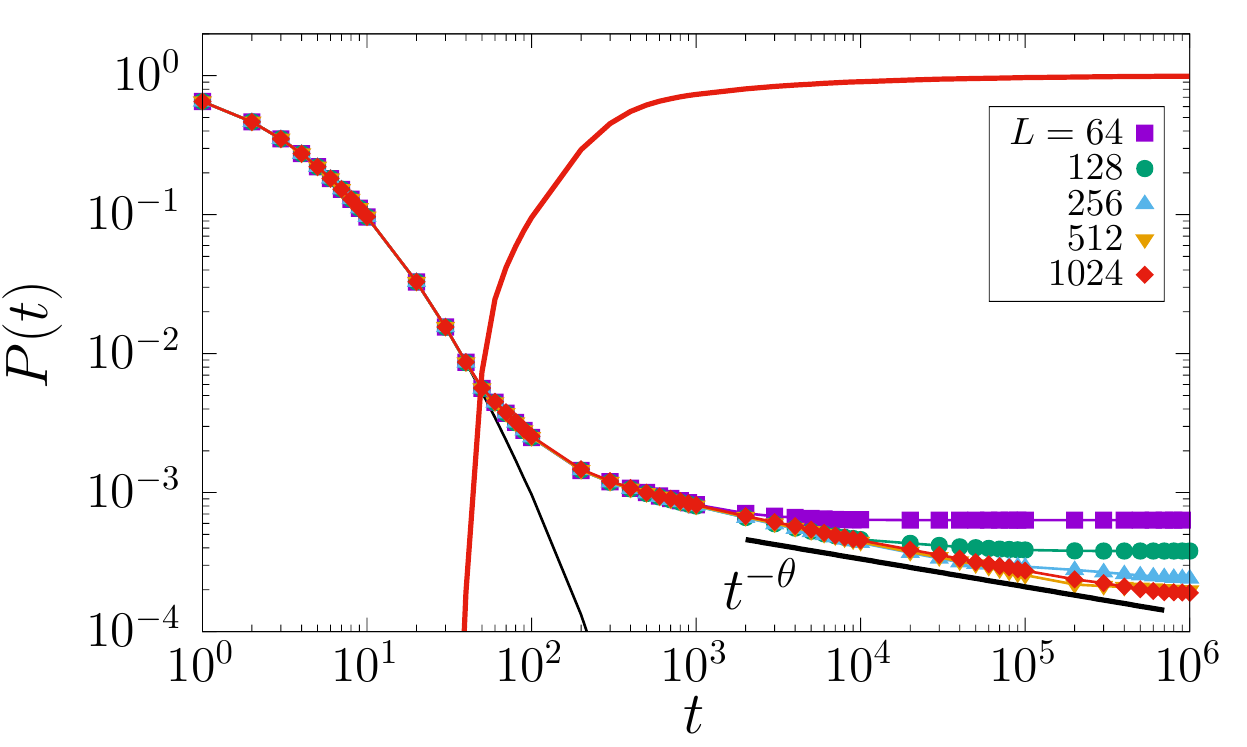}
    \caption{Persistence $P(t)$ for increasing linear sizes $L$, $\Delta\eta=10^{-2}$ and $\gamma\to\infty$. In the initial regime there are no zealots and the behavior follows the VM ($L=64$, thin black line). As the density of zealots increases (red thick line), $P(t)$ departs from this behavior and eventually develops a power-law, $P(t) \sim t^{-\theta}$. The thick black line shows the 2d Ising behavior whose exponent, after a quench from high temperature, is $\theta\simeq 0.2$~\cite{DeBrGo94,Stauffer94,BlCuPi14}.}
\label{fig.persistence}
\end{figure}

Fig.~\ref{fig.persistence} presents the behavior of the persistence $P(t)$. Similarly to the 1d case, the initial regime is equivalent to the VM~\cite{BeFrKr96,HoGo98} as zealots are still absent.
However, upon the sudden rise in the number of zealots (red solid line), $P(t)$ slows down and deviates from the VM curve (thin black line).
For increasing system sizes, $P(t)$ develops a power law behavior (thick black line), $P(t)\sim t^{-\theta}$, whose exponent is consistent with the one for the 2d Ising model after a temperature quench from high temperature, $\theta\simeq 0.2$~\cite{DeBrGo94,Stauffer94,BlCuPi14}.
Most of the persistent spins are in the zealots bulk region, and the fluctuating interface between different opinions must collide with the internal interface in order to destabilize the zealots, originating the slowing down.

\begin{figure}[ht]
    \includegraphics[width=\columnwidth]{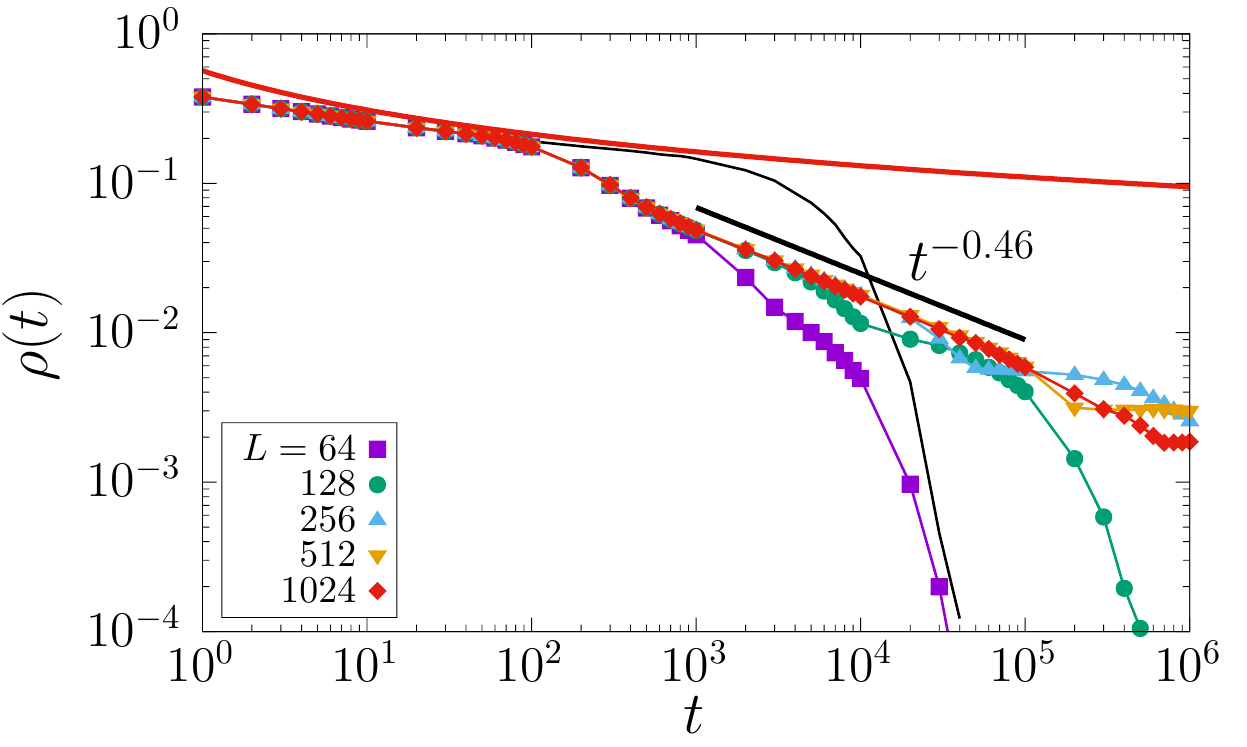}
    \caption{Density of active interfaces $\rho(t)$, defined as the fraction of neighboring agents with different opinions, for $\Delta\eta=10^{-2}$, $\gamma\to\infty$ and different sizes. The VM results for $L=64$ are shown for comparison (thin black line) along with the expected, asymptotic behavior~\cite{FrKr96,TaCuPi15} (thick red line). Deviations from the VM behavior start again when zealots rapidly invade the lattice. Since domains become smoother, the total perimeter, measured by $\rho(t)$, presents a strong decrease relative to VM. For large enough systems, a power-law with an exponent slightly below 1/2, $\rho(t) \sim t^{-0.46}$, develops.}
\label{fig.conec}
\end{figure}

Fig.~\ref{fig.conec} shows the density $\rho(t)$ of sites that have a nearest neighbor with a different opinion, i.e., located on the rough active interface that separates two domains.
For $\gamma\to\infty$, sites belonging to a stripe whose width corresponds to the mean height of the surface will have a high probability of having small values of $\eta_i$, thus following the VM dynamics.
After zealots are formed, $\rho(t)$ presents a strong decrease, deviating from the slow inverse logarithm behavior of the VM~\cite{FrKr96}.
Moreover, for large enough systems, a power-law develops, $\rho(t)\sim t^{-0.46}$, whose exponent is consistent with the model A universality class, albeit slightly below 1/2. 
This deviation from the characteristic 1/2 exponent of the curvature driven coarsening was observed in similar models~\cite{CaEgSa06,DaGa08,VeVa18,MuBiPa20}.
Notice that although larger clusters have an underlying structural frame provided by the zealots bulk, smaller domains that are formed by fragmentation close to the surface are, in general, purely VM and contribute with a slower, logarithmic time dependence.
This effect, along with those samples that have a longer lived metastable structure (see below) seems to be the main mechanism explaining this small exponent difference.
Despite its short-range roughness analogous to the VM, the smaller value of $\rho(t)$ shown in Fig.~\ref{fig.conec} indicates that on larger scale the interfaces are smoother.
This is a consequence of the zealots bulk that, growing by a curvature-driven dynamics, have a strong influence on the long range properties of the VM stripe, providing a rather smooth support that decreases the overall perimeter.

\begin{figure}[ht]
\includegraphics[width=\columnwidth]{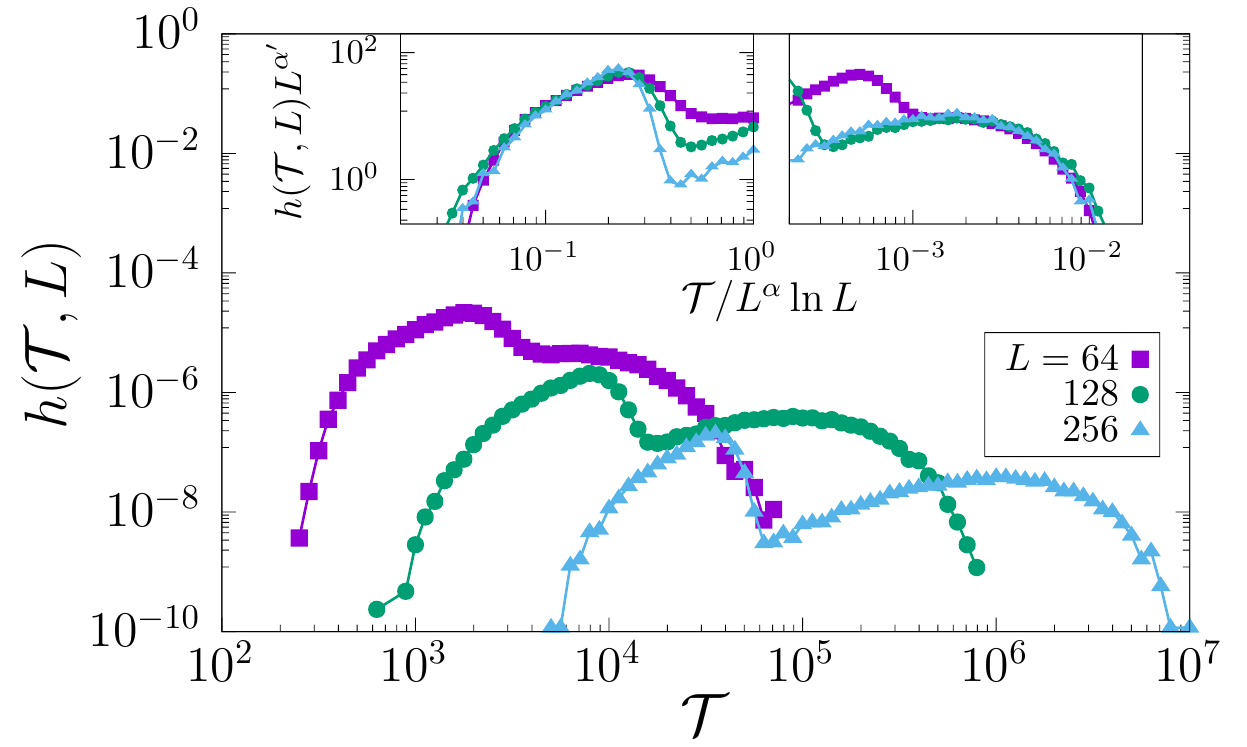}
\caption{Distribution $h({\cal T},L)$ of the consensus time ${\cal T}$ with $\Delta\eta=10^{-2}$ and $\gamma\to\infty$. Our model, differently from the VM where there is a single timescale~\cite{TaCuPi15}, develops also a longer timescale related to transient stripes, similar to the IM0. Good collapses are obtained by rescaling the horizontal axis with $L^\alpha \ln L$: $\alpha \simeq 2$ for the first peak (left inset) e $\alpha \simeq 3.5$ for the second one (right inset). In the vertical direction, $\alpha'=3.5$ for both peaks.}
\label{fig.hist}
\end{figure}

In the IM0, the asymptotic state is either fully magnetized or divided into multiple (most often two) parallel stripes~\cite{DeOlSt96,Lipowski99,SpKrRe01a,SpKrRe01b,BaKrRe09,OlKrRe12,GoPl18} and either of these possible fates are decided earlier in the dynamics, when it approaches the percolative critical point~\cite{ArBrCuSi07,SiArBrCu07,BlCoCuPi14,BlCuPiTa17,AzAlOlAr22}. 
The time to attain the former grows as ${\cal T}\sim L^2$ for most of the initial states~\cite{Lipowski99}. However, a small fraction of these initial states develops diagonal stripes that slow down the dynamics and whose characteristic time increases as ${\cal T}\sim L^{3.5}$~\cite{Lipowski99,SpKrRe01a,GoPl18}. For the VM, since it lacks surface tension and straight interfaces are unstable, no structures resembling stripes are formed. All initial states do converge to consensus in a time whose average scales as $\langle{\cal T}\rangle\sim L^2\ln L$~\cite{Krapivsky92,Liggett99,TaCuPi15}. As a consequence, the single-peaked consensus time distribution, $h({\cal T},L)$, obeys the scaling relation $h({\cal T},L)=L^{-\alpha'}H({\cal T}/L^{\alpha}\ln L)$~\cite{TaCuPi15}. In our model, once zealots are formed and curvature driven  dynamics becomes important, some initial states, with clusters that wrap the system along a single direction, develop transient structures that are similar to stripes. Nonetheless, since the dynamics at the surface is driven by interfacial noise, these stripes remain unstable and the system eventually converge to consensus. Nonetheless, the presence of such stripes introduces a new timescale to attain consensus.  This is the origin of the second peak in Fig.~\ref{fig.hist} for the distribution $h({\cal T},L)$ and has been observed in similar models~\cite{CaEgSa06,CaBaLo09,VoRe12,VeVa18,MuBiPa20}. The main contribution, however, comes from those initial states that are not delayed and consensus is attained faster because, early in the dynamics, a percolating cluster, wrapping the system in two directions~\cite{TaCuPi15}, has been formed. Notice that although a double peak structure was also observed for the IM0~\cite{GoPl18}, in our model the second peak also has the contribution of those states whose stripes are parallel to the lattice directions, while in the IM0 they lead to absorbing states and only diagonally striped configurations do contribute. 
In the (left) inset of Fig.~\ref{fig.hist} an excellent collapse is obtained with $L^\alpha \ln L$ and $\alpha \simeq 2$. This is similar to the $L^2$ scaling of the corresponding first peak for the IM0 but also includes the logarithmic correction from the VM~\cite{TaCuPi15}. 
The second peak collapse, with $\alpha\simeq 3.5$ is shown in the right inset.

\begin{figure}[ht]
\includegraphics[width=\columnwidth]{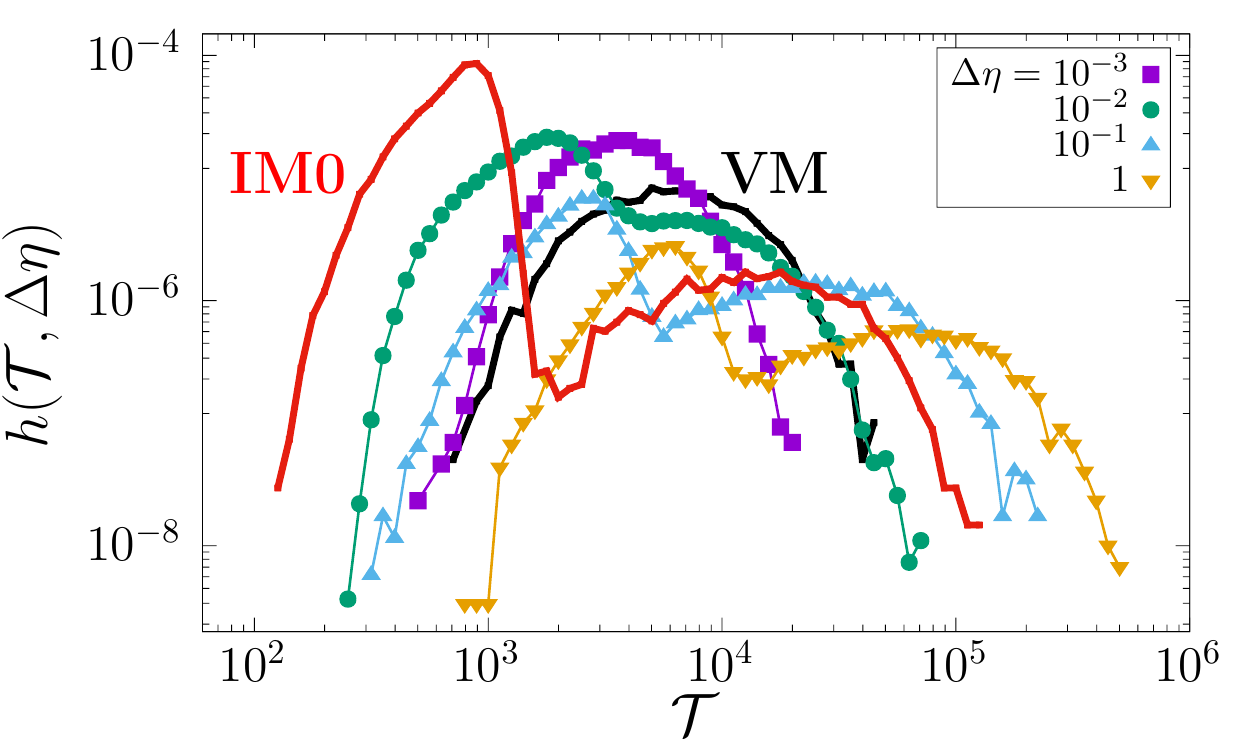}
\caption{Consensus time distribution $h({   \cal T},L)$ for a single size, $L=64$, and different values of $\Delta\eta$. For comparison, the VM (black line) and the IM0 (red line) are also shown. Averages over $10^4$ samples were considered.
}
\label{fig.histodetas}
\end{figure}

Although one could expect a simple interpolation, by varying $\Delta\eta$, from the VM (when $\Delta\eta\to 0$) and the IM0 (for $\Delta\eta\to 1$), Fig.~\ref{fig.histodetas} shows that the dependence is non-trivial. 
For $\Delta\eta=10^{-3}$, there is a single, large peak of $h({\cal T},L)$. 
Increasing it slightly, a second peak appears while the first one moves towards the position of the first peak of the IM0.
But after attaining a minimum value, this tendency is reversed for intermediate values of $\Delta\eta$ and start approaching the VM peak.
Notice that, for all values of $\Delta\eta$, the first peak is within the first peak of the IM0 and the VM single peak.
For $\Delta\eta=1$, however, the whole distribution has the largest displacement away from the IM0 distribution.
Thus,  the active region on the surface of all clusters, whatever its width, always have a delaying effect. 
The second, smaller peak, once formed, does not seem to present a minimum and always move to larger times, becoming even larger (and wider) than the corresponding peak of the IM0.

\begin{figure}[htb]
\includegraphics[width=\columnwidth]{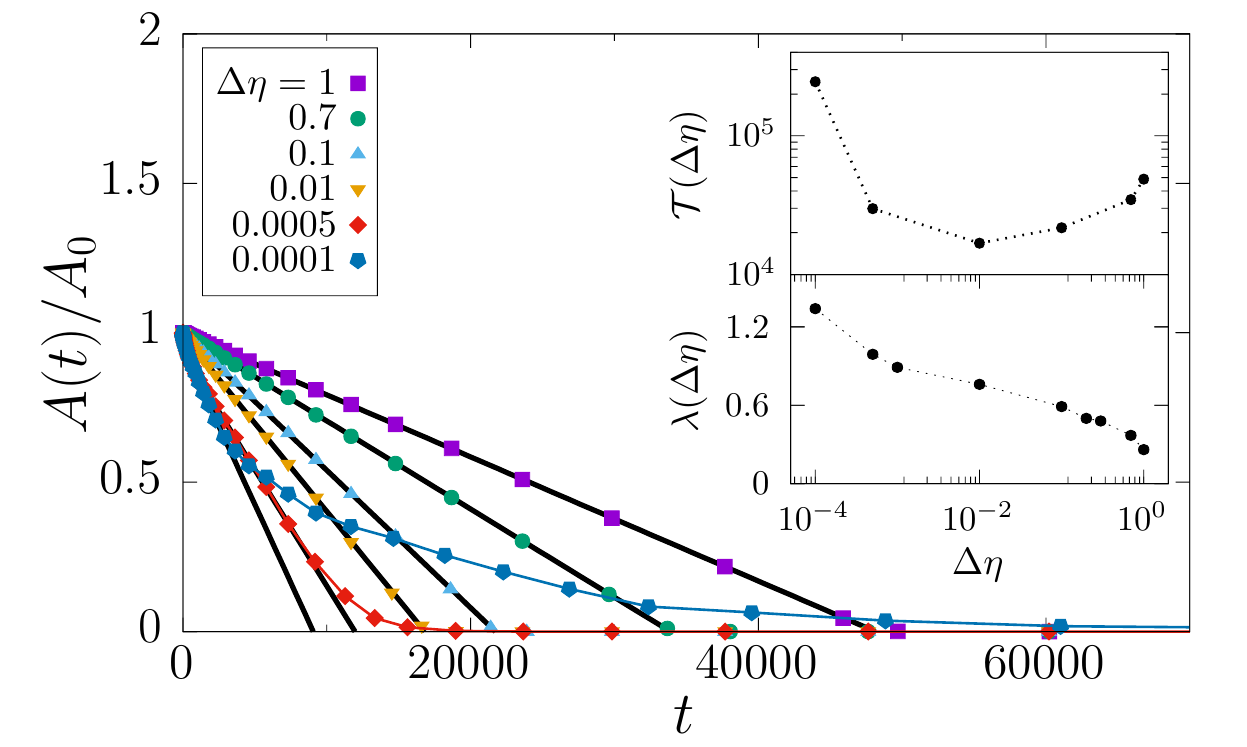}
\caption{Time evolution of the average area $A(t)/A_0$ of an initially circular domain containing a single opinion (Fig.~\ref{fig.snap01}, bottom row), for $L=256$ and several values of $\Delta\eta$. Averages are over 1000 samples. The straight black lines are linear fits at short times, whose declivity $\lambda$ is the rate with which the area shrinks (bottom inset). Notice that above the minimum  for $\Delta\eta\gtrsim\Delta\eta_{\text{min}}\approx 10^{-2}$, the linear behavior, characteristic of the Ising model~\cite{GrGu83,SaGrSa83}, persists during the whole interval, while for smaller values, it crosses over to the logarithmic behavior. The average consensus time ${\cal T}(\Delta\eta)$ is shown in the top inset. 
} 
\label{fig.circulo}
\end{figure}

In order to understand the origin of the above minimum, we consider a specially prepared initial state that prevents the formation of stripes.
As illustrated in the bottom row of Fig.~\ref{fig.snap01}, one opinion is initially fully embedded in a circular domain while the other one surrounds it~\cite{DaCa07,DaGa08,DoSrSzKo16}.
Fig.~\ref{fig.circulo} shows the time evolution of the relative area of the selected opinion $A(t)/A(0)$, regardless of the fragmentation that may occur on the surface.
For very small values of $\Delta\eta$, the possible presence of zealots late in the dynamics has little impact and the behavior is logarithmic, following the VM.
As $\Delta\eta$ increases, the zealots bulk gets formed along with the VM  region closer to the surface whose width depends on $\Delta\eta$.
While the drop and its fragments decrease in size, the VM region roughly keeps its width, forcing the internal border with the zealots bulk to recede, being the first region to disappear.
During the time interval while zealots are present, the behavior is linear, similar to the IM0, $A(t)\simeq A_0 (1-\lambda t)$.
The parameter $\lambda$ is a monotonously decreasing function of $\Delta\eta$, bottom inset of Fig.~\ref{fig.circulo}.
As $\Delta\eta$ increases from $10^{-4}$ to 1, $\lambda$ decreases from, roughly, 1.4 to 0.2, while for the IM0, $\lambda\simeq 2$~\cite{ArCuPi15} and the drop disappears faster than all cases considered here. 
An interesting consequence is that the faster the initial linear decrease is ($\Delta\eta\to 0$), the sooner the behavior of $A(t$) deviates from it.
Moreover, for small $\Delta\eta$, once $A(t)$ becomes logarithmic, the average consensus time increases.
On the other hand, for $\Delta\eta\to 1$, the VM region is very small and, because most of the agents are zealots, the dynamics slows down and $\lambda$ is small.
In this case, even if the deviations from the linear behavior cannot be seen in the linear scale of Fig.~\ref{fig.circulo}, the average consensus time is large once again.
Thus, $\langle{\cal T}\rangle$ has a minimum~\cite{StTeSc08,StTeSc08b,WaLiWaZhWa14} at an intermediate value, close to $\Delta\eta\simeq 10^{-2}$. Notice that although the above behavior is rather clear for a single droplet, once a more general, random initial state is considered the trend for the average area is not (not shown). The probable origin is the VM region that dresses each compact bulk. 
Being easily fragmented, there is a large contribution of small domains to the average from sizes that are much less frequent in the IM0.

\section{Conclusions}
\label{sec.conclusions}

We introduced an opinion model whose agents have intermediate levels of confidence that may interfere in their process of changing opinion.
While all agents have low confidence, in the beginning of the dynamics, the model is equivalent to the VM, where a single contact with a different opinion is enough for an agent to change its own.
Opinion reinforcement builds up confidence and, above a certain threshold, the agent becomes a kinetically constrained zealot whose opinion is frozen.
But regardless the nature of the agent, the variables characterizing their confidences keep evolving as the agents interact with their neighborhood.
We considered two limiting cases depending on the parameter $\gamma$ that rescales the confidence after a confront with a different opinion.
For $\gamma=1$, the zealot state is irreversible while for $\gamma\to\infty$ the dynamics allows the zealot state to be reversed and the coexistence with normal voters.
Similar opinions segregate in spatial domains and, in their bulk, because of the positive reinforcement, zealots first appear once the confidence threshold $\phi$ is attained. 
The zealots being frozen, there is an increased probability of repeated contacts with their non-zealots neighbors, increasing their confidence as well.
This mechanism of noise reduction smooths the internal interface between zealots and normal voters, both with the same opinion.
This smoother underneath surface induces an effective surface tension, acting as a structural frame that turns the dynamics from interfacial noise to curvature-driven.
As a consequence, several properties become analogous to those of the low temperature Ising model with non-conserved order parameter, in the Allen-Cahn (Model A) universality class~\cite{Bray94}.
Normal voters, instead, are confined close to the surface because the constant interaction with the opposite opinion that keeps their confidence low and their amount depends on $\Delta\eta$, i.e., how fast the zealot bulk grows.
Although the curvature-driven growth has been already observed in other variations of the VM~\cite{CaEgSa06,DaCa07,DaGa08,CaBaLo09,VoRe12,ZhLiKoSz14,DoSrSzKo16,RoSe17,VeVa18,MuBiPa20}, our model allows to tune the amount of normal voters close to the surface through $\Delta\eta$.
Interestingly, the internal surface induces a curvature-driven growth but the actual surface is driven by the interface noise typical of the VM, becoming rougher and more fragmented than other models.

The width of the normal voters region close to the surface, that depends on $\Delta\eta$ (see Fig.~\ref{fig.wall}), determines how fast the domains shrink and, consequently, the exit time, i.e., how long does it take to attain consensus. 
While a large domain shrinks, the zealot bulk disappears first and that cluster dynamics is no longer driven by the curvature of the interface. This late regime is dominated by the normal voters and the dynamics becomes logarithmic. How important will be this final regime depends on the width of the stripe with normal voters around the bulk. 
When $\Delta\eta$ is large, it is thin and most of the dynamics is dominated by the curvature-driven mechanism induced by the zealots. However, despite being curvature-driven, the dynamics is slow because of the large amount of zealots, it is necessary first to turn them into normal agents.
On the other hand, for small $\Delta\eta$, the stripe is thick and once the bulk disappears leaving only the normal voters, there is a crossover to a slower, logarithmic regime.
It is the interplay between both mechanisms that explain the existence of a minimum consensus time as a function of $\Delta\eta$, directly related to the superficial normal voters.

There are some possible generalizations of the model that would be interesting to investigate. For example, individual heterogeneities~\cite{MaGiRe10} in the values of $\gamma$, $\{\gamma_i\}$, can be considered. If some of the agents have $\gamma_i\leq 1$, they may become permanent zealots. Another possible modification is to reintroduce the conservation of the magnetization, that is present in the pure VM but broken in our model, by a local conservation rule~\cite{CaDaGaRo13}. Such conserved order parameter dynamics is known to be in a different universality class from the non-conserved case~\cite{Bray94,SiSaArNrCu09} and its effect in the present model are worth investigating. 


\begin{thebibliography}{62}
\expandafter\ifx\csname natexlab\endcsname\relax\def\natexlab#1{#1}\fi
\expandafter\ifx\csname bibnamefont\endcsname\relax
  \def\bibnamefont#1{#1}\fi
\expandafter\ifx\csname bibfnamefont\endcsname\relax
  \def\bibfnamefont#1{#1}\fi
\expandafter\ifx\csname citenamefont\endcsname\relax
  \def\citenamefont#1{#1}\fi
\expandafter\ifx\csname url\endcsname\relax
  \def\url#1{\texttt{#1}}\fi
\expandafter\ifx\csname urlprefix\endcsname\relax\def\urlprefix{URL }\fi
\providecommand{\bibinfo}[2]{#2}
\providecommand{\eprint}[2][]{\url{#2}}

\bibitem[{\citenamefont{Castellano et~al.}(2009)\citenamefont{Castellano,
  Fortunato, and Loreto}}]{CaFoLo09}
\bibinfo{author}{\bibfnamefont{C.}~\bibnamefont{Castellano}},
  \bibinfo{author}{\bibfnamefont{S.}~\bibnamefont{Fortunato}},
  \bibnamefont{and} \bibinfo{author}{\bibfnamefont{V.}~\bibnamefont{Loreto}},
  \bibinfo{journal}{Rev. Mod. Phys.} \textbf{\bibinfo{volume}{81}},
  \bibinfo{pages}{591} (\bibinfo{year}{2009}).

\bibitem[{\citenamefont{Baronchelli}(2018)}]{Baronchelli18}
\bibinfo{author}{\bibfnamefont{A.}~\bibnamefont{Baronchelli}},
  \bibinfo{journal}{R. Soc. open sci.} \textbf{\bibinfo{volume}{5}},
  \bibinfo{pages}{172189} (\bibinfo{year}{2018}).

\bibitem[{\citenamefont{Redner}(2019)}]{Redner19}
\bibinfo{author}{\bibfnamefont{S.}~\bibnamefont{Redner}}, \bibinfo{journal}{C.
  R. Physique} \textbf{\bibinfo{volume}{20}}, \bibinfo{pages}{275}
  (\bibinfo{year}{2019}).

\bibitem[{\citenamefont{Lynas et~al.}(2021)\citenamefont{Lynas, Houlton, and
  Perry}}]{LyHoPe21}
\bibinfo{author}{\bibfnamefont{M.}~\bibnamefont{Lynas}},
  \bibinfo{author}{\bibfnamefont{B.~Z.} \bibnamefont{Houlton}},
  \bibnamefont{and} \bibinfo{author}{\bibfnamefont{S.}~\bibnamefont{Perry}},
  \bibinfo{journal}{Environ. Res. Lett.} \textbf{\bibinfo{volume}{16}},
  \bibinfo{pages}{114005} (\bibinfo{year}{2021}).

\bibitem[{\citenamefont{Sturgis et~al.}(2021)\citenamefont{Sturgis,
  Brunton-Smith, and Jackson}}]{StBrJa21}
\bibinfo{author}{\bibfnamefont{P.}~\bibnamefont{Sturgis}},
  \bibinfo{author}{\bibfnamefont{I.}~\bibnamefont{Brunton-Smith}},
  \bibnamefont{and} \bibinfo{author}{\bibfnamefont{J.}~\bibnamefont{Jackson}},
  \bibinfo{journal}{Nat. Hum. Behav.} \textbf{\bibinfo{volume}{5}},
  \bibinfo{pages}{1528} (\bibinfo{year}{2021}).

\bibitem[{\citenamefont{Krapivsky}(1992)}]{Krapivsky92}
\bibinfo{author}{\bibfnamefont{P.~L.} \bibnamefont{Krapivsky}},
  \bibinfo{journal}{Phys. Rev. A} \textbf{\bibinfo{volume}{45}},
  \bibinfo{pages}{1067} (\bibinfo{year}{1992}).

\bibitem[{\citenamefont{Liggett}(1999)}]{Liggett99}
\bibinfo{author}{\bibfnamefont{T.~M.} \bibnamefont{Liggett}},
  \emph{\bibinfo{title}{{Stochastic Interacting Systems: Contact, Voter and
  Exclusion Processes}}} (\bibinfo{publisher}{Springer}, \bibinfo{year}{1999}).

\bibitem[{\citenamefont{Frachebourg and Krapivsky}(1996)}]{FrKr96}
\bibinfo{author}{\bibfnamefont{L.}~\bibnamefont{Frachebourg}} \bibnamefont{and}
  \bibinfo{author}{\bibfnamefont{P.~L.} \bibnamefont{Krapivsky}},
  \bibinfo{journal}{Phys. Rev. E} \textbf{\bibinfo{volume}{53}},
  \bibinfo{pages}{R3009} (\bibinfo{year}{1996}).

\bibitem[{\citenamefont{Ben-Naim et~al.}(1996)\citenamefont{Ben-Naim,
  Frachebourg, and Krapivsky}}]{BeFrKr96}
\bibinfo{author}{\bibfnamefont{E.}~\bibnamefont{Ben-Naim}},
  \bibinfo{author}{\bibfnamefont{L.}~\bibnamefont{Frachebourg}},
  \bibnamefont{and} \bibinfo{author}{\bibfnamefont{P.~L.}
  \bibnamefont{Krapivsky}}, \bibinfo{journal}{Phys. Rev. E}
  \textbf{\bibinfo{volume}{53}}, \bibinfo{pages}{3078} (\bibinfo{year}{1996}).

\bibitem[{\citenamefont{Dornic et~al.}(2001)\citenamefont{Dornic, Chaté,
  Chave, and Hinrichsen}}]{DoChChHi01}
\bibinfo{author}{\bibfnamefont{I.}~\bibnamefont{Dornic}},
  \bibinfo{author}{\bibfnamefont{H.}~\bibnamefont{Chaté}},
  \bibinfo{author}{\bibfnamefont{J.}~\bibnamefont{Chave}}, \bibnamefont{and}
  \bibinfo{author}{\bibfnamefont{H.}~\bibnamefont{Hinrichsen}},
  \bibinfo{journal}{Phys. Rev. Lett.} \textbf{\bibinfo{volume}{87}},
  \bibinfo{pages}{045701} (\bibinfo{year}{2001}).

\bibitem[{\citenamefont{Tartaglia et~al.}(2015)\citenamefont{Tartaglia,
  Cugliandolo, and Picco}}]{TaCuPi15}
\bibinfo{author}{\bibfnamefont{A.}~\bibnamefont{Tartaglia}},
  \bibinfo{author}{\bibfnamefont{L.~F.} \bibnamefont{Cugliandolo}},
  \bibnamefont{and} \bibinfo{author}{\bibfnamefont{M.}~\bibnamefont{Picco}},
  \bibinfo{journal}{Phys. Rev. E} \textbf{\bibinfo{volume}{92}},
  \bibinfo{pages}{042109} (\bibinfo{year}{2015}).

\bibitem[{\citenamefont{Mobilia}(2003)}]{Mobilia03}
\bibinfo{author}{\bibfnamefont{M.}~\bibnamefont{Mobilia}},
  \bibinfo{journal}{Phys. Rev. Lett.} \textbf{\bibinfo{volume}{91}},
  \bibinfo{pages}{028701} (\bibinfo{year}{2003}).

\bibitem[{\citenamefont{Mobilia and Georgiev}(2005)}]{MoGe05}
\bibinfo{author}{\bibfnamefont{M.}~\bibnamefont{Mobilia}} \bibnamefont{and}
  \bibinfo{author}{\bibfnamefont{I.~T.} \bibnamefont{Georgiev}},
  \bibinfo{journal}{Phys. Rev. E} \textbf{\bibinfo{volume}{71}},
  \bibinfo{pages}{046102} (\bibinfo{year}{2005}).

\bibitem[{\citenamefont{Mobilia et~al.}(2007)\citenamefont{Mobilia, Petersen,
  and Redner}}]{MoPeRe07}
\bibinfo{author}{\bibfnamefont{M.}~\bibnamefont{Mobilia}},
  \bibinfo{author}{\bibfnamefont{A.}~\bibnamefont{Petersen}}, \bibnamefont{and}
  \bibinfo{author}{\bibfnamefont{S.}~\bibnamefont{Redner}},
  \bibinfo{journal}{J. Stat. Mech.} p. \bibinfo{pages}{P08029}
  (\bibinfo{year}{2007}).

\bibitem[{\citenamefont{Galam and Jacobs}(2007)}]{GaJa07}
\bibinfo{author}{\bibfnamefont{S.}~\bibnamefont{Galam}} \bibnamefont{and}
  \bibinfo{author}{\bibfnamefont{F.}~\bibnamefont{Jacobs}},
  \bibinfo{journal}{Physica A} \textbf{\bibinfo{volume}{381}},
  \bibinfo{pages}{366} (\bibinfo{year}{2007}).

\bibitem[{\citenamefont{Colaiori and Castellano}(2016)}]{CoCa16}
\bibinfo{author}{\bibfnamefont{F.}~\bibnamefont{Colaiori}} \bibnamefont{and}
  \bibinfo{author}{\bibfnamefont{C.}~\bibnamefont{Castellano}},
  \bibinfo{journal}{J. Stat. Mech.:} p. \bibinfo{pages}{033401}
  (\bibinfo{year}{2016}).

\bibitem[{\citenamefont{Dall'Asta and Castellano}(2007)}]{DaCa07}
\bibinfo{author}{\bibfnamefont{L.}~\bibnamefont{Dall'Asta}} \bibnamefont{and}
  \bibinfo{author}{\bibfnamefont{C.}~\bibnamefont{Castellano}},
  \bibinfo{journal}{EPL} \textbf{\bibinfo{volume}{77}}, \bibinfo{pages}{60005}
  (\bibinfo{year}{2007}).

\bibitem[{\citenamefont{Stark et~al.}(2008{\natexlab{a}})\citenamefont{Stark,
  Tessone, and Schweitzer}}]{StTeSc08}
\bibinfo{author}{\bibfnamefont{H.-U.} \bibnamefont{Stark}},
  \bibinfo{author}{\bibfnamefont{C.~J.} \bibnamefont{Tessone}},
  \bibnamefont{and}
  \bibinfo{author}{\bibfnamefont{F.}~\bibnamefont{Schweitzer}},
  \bibinfo{journal}{Phys. Rev. Lett.} \textbf{\bibinfo{volume}{101}},
  \bibinfo{pages}{018701} (\bibinfo{year}{2008}{\natexlab{a}}).

\bibitem[{\citenamefont{Stark et~al.}(2008{\natexlab{b}})\citenamefont{Stark,
  Tessone, and Schweitzer}}]{StTeSc08b}
\bibinfo{author}{\bibfnamefont{H.-U.} \bibnamefont{Stark}},
  \bibinfo{author}{\bibfnamefont{C.~J.} \bibnamefont{Tessone}},
  \bibnamefont{and}
  \bibinfo{author}{\bibfnamefont{F.}~\bibnamefont{Schweitzer}},
  \bibinfo{journal}{Adv. Compl. Syst.} \textbf{\bibinfo{volume}{11}},
  \bibinfo{pages}{551} (\bibinfo{year}{2008}{\natexlab{b}}).

\bibitem[{\citenamefont{Dall’Asta and Galla}(2008)}]{DaGa08}
\bibinfo{author}{\bibfnamefont{L.}~\bibnamefont{Dall’Asta}} \bibnamefont{and}
  \bibinfo{author}{\bibfnamefont{T.}~\bibnamefont{Galla}}, \bibinfo{journal}{J.
  Phys. A: Math. Gen.} \textbf{\bibinfo{volume}{41}}, \bibinfo{pages}{435003}
  (\bibinfo{year}{2008}).

\bibitem[{\citenamefont{Lambiotte et~al.}(2009)\citenamefont{Lambiotte,
  Saramäki, and Blondel}}]{LaSaBl09}
\bibinfo{author}{\bibfnamefont{R.}~\bibnamefont{Lambiotte}},
  \bibinfo{author}{\bibfnamefont{J.}~\bibnamefont{Saramäki}},
  \bibnamefont{and} \bibinfo{author}{\bibfnamefont{V.~D.}
  \bibnamefont{Blondel}}, \bibinfo{journal}{Phys. Rev. E}
  \textbf{\bibinfo{volume}{79}}, \bibinfo{pages}{046107}
  (\bibinfo{year}{2009}).

\bibitem[{\citenamefont{Volovik and Redner}(2012)}]{VoRe12}
\bibinfo{author}{\bibfnamefont{D.}~\bibnamefont{Volovik}} \bibnamefont{and}
  \bibinfo{author}{\bibfnamefont{S.}~\bibnamefont{Redner}},
  \bibinfo{journal}{J. Stat. Mech.} p. \bibinfo{pages}{P04003}
  (\bibinfo{year}{2012}).

\bibitem[{\citenamefont{Martins and Galam}(2013)}]{MaGa13}
\bibinfo{author}{\bibfnamefont{A.~C.~R.} \bibnamefont{Martins}}
  \bibnamefont{and} \bibinfo{author}{\bibfnamefont{S.}~\bibnamefont{Galam}},
  \bibinfo{journal}{Phys. Rev. E} \textbf{\bibinfo{volume}{87}},
  \bibinfo{pages}{042807} (\bibinfo{year}{2013}).

\bibitem[{\citenamefont{Wang et~al.}(2014)\citenamefont{Wang, Liu, Wang, Zhang,
  and Wang}}]{WaLiWaZhWa14}
\bibinfo{author}{\bibfnamefont{Z.}~\bibnamefont{Wang}},
  \bibinfo{author}{\bibfnamefont{Y.}~\bibnamefont{Liu}},
  \bibinfo{author}{\bibfnamefont{L.}~\bibnamefont{Wang}},
  \bibinfo{author}{\bibfnamefont{Y.}~\bibnamefont{Zhang}}, \bibnamefont{and}
  \bibinfo{author}{\bibfnamefont{Z.}~\bibnamefont{Wang}},
  \bibinfo{journal}{Sci. Rep.} \textbf{\bibinfo{volume}{4}},
  \bibinfo{pages}{3597} (\bibinfo{year}{2014}).

\bibitem[{\citenamefont{Brugna and Toscani}(2015)}]{BrTo15}
\bibinfo{author}{\bibfnamefont{C.}~\bibnamefont{Brugna}} \bibnamefont{and}
  \bibinfo{author}{\bibfnamefont{G.}~\bibnamefont{Toscani}},
  \bibinfo{journal}{Phys. Rev. E} \textbf{\bibinfo{volume}{92}},
  \bibinfo{pages}{052818} (\bibinfo{year}{2015}).

\bibitem[{\citenamefont{Velásquez-Rojas and Vazquez}(2018)}]{VeVa18}
\bibinfo{author}{\bibfnamefont{F.}~\bibnamefont{Velásquez-Rojas}}
  \bibnamefont{and} \bibinfo{author}{\bibfnamefont{F.}~\bibnamefont{Vazquez}},
  \bibinfo{journal}{J. Stat. Mech.} p. \bibinfo{pages}{043403}
  (\bibinfo{year}{2018}).

\bibitem[{\citenamefont{Castelló et~al.}(2006)\citenamefont{Castelló,
  Eguíluz, and San~Miguel}}]{CaEgSa06}
\bibinfo{author}{\bibfnamefont{X.}~\bibnamefont{Castelló}},
  \bibinfo{author}{\bibfnamefont{V.~M.} \bibnamefont{Eguíluz}},
  \bibnamefont{and}
  \bibinfo{author}{\bibfnamefont{M.}~\bibnamefont{San~Miguel}},
  \bibinfo{journal}{New J. Phys.} \textbf{\bibinfo{volume}{8}},
  \bibinfo{pages}{308} (\bibinfo{year}{2006}).

\bibitem[{\citenamefont{Castelló et~al.}(2009)\citenamefont{Castelló,
  Baronchelli, and Loreto}}]{CaBaLo09}
\bibinfo{author}{\bibfnamefont{X.}~\bibnamefont{Castelló}},
  \bibinfo{author}{\bibfnamefont{A.}~\bibnamefont{Baronchelli}},
  \bibnamefont{and} \bibinfo{author}{\bibfnamefont{V.}~\bibnamefont{Loreto}},
  \bibinfo{journal}{Eur. Phys. J. B} \textbf{\bibinfo{volume}{71}},
  \bibinfo{pages}{557} (\bibinfo{year}{2009}).

\bibitem[{\citenamefont{Zhang et~al.}(2014)\citenamefont{Zhang, Lim, Korniss,
  and Szymanski}}]{ZhLiKoSz14}
\bibinfo{author}{\bibfnamefont{W.}~\bibnamefont{Zhang}},
  \bibinfo{author}{\bibfnamefont{C.~C.} \bibnamefont{Lim}},
  \bibinfo{author}{\bibfnamefont{G.}~\bibnamefont{Korniss}}, \bibnamefont{and}
  \bibinfo{author}{\bibfnamefont{B.~K.} \bibnamefont{Szymanski}},
  \bibinfo{journal}{Sci. Rep.} \textbf{\bibinfo{volume}{4}},
  \bibinfo{pages}{5568} (\bibinfo{year}{2014}).

\bibitem[{\citenamefont{Doyle et~al.}(2016)\citenamefont{Doyle, Sreenivasan,
  Szymanski, and Korniss}}]{DoSrSzKo16}
\bibinfo{author}{\bibfnamefont{C.}~\bibnamefont{Doyle}},
  \bibinfo{author}{\bibfnamefont{S.}~\bibnamefont{Sreenivasan}},
  \bibinfo{author}{\bibfnamefont{B.~K.} \bibnamefont{Szymanski}},
  \bibnamefont{and} \bibinfo{author}{\bibfnamefont{G.}~\bibnamefont{Korniss}},
  \bibinfo{journal}{Physica A} \textbf{\bibinfo{volume}{443}},
  \bibinfo{pages}{316} (\bibinfo{year}{2016}).

\bibitem[{\citenamefont{Roy and Sen}(2017)}]{RoSe17}
\bibinfo{author}{\bibfnamefont{P.}~\bibnamefont{Roy}} \bibnamefont{and}
  \bibinfo{author}{\bibfnamefont{P.}~\bibnamefont{Sen}},
  \bibinfo{journal}{Phys. Rev. E} \textbf{\bibinfo{volume}{95}},
  \bibinfo{pages}{020101(R)} (\bibinfo{year}{2017}).

\bibitem[{\citenamefont{Mukherjee et~al.}(2020)\citenamefont{Mukherjee, Biswas,
  and Sen}}]{MuBiPa20}
\bibinfo{author}{\bibfnamefont{S.}~\bibnamefont{Mukherjee}},
  \bibinfo{author}{\bibfnamefont{S.}~\bibnamefont{Biswas}}, \bibnamefont{and}
  \bibinfo{author}{\bibfnamefont{P.}~\bibnamefont{Sen}},
  \bibinfo{journal}{Phys. Rev. E} \textbf{\bibinfo{volume}{102}},
  \bibinfo{pages}{012316} (\bibinfo{year}{2020}).

\bibitem[{\citenamefont{Bray}(1994)}]{Bray94}
\bibinfo{author}{\bibfnamefont{A.~J.} \bibnamefont{Bray}},
  \bibinfo{journal}{Adv. Phys.} \textbf{\bibinfo{volume}{43}},
  \bibinfo{pages}{481} (\bibinfo{year}{1994}).

\bibitem[{\citenamefont{Bray et~al.}(2013)\citenamefont{Bray, Majumdar, and
  Schehr}}]{BrMaSc13}
\bibinfo{author}{\bibfnamefont{A.~J.} \bibnamefont{Bray}},
  \bibinfo{author}{\bibfnamefont{S.~N.} \bibnamefont{Majumdar}},
  \bibnamefont{and} \bibinfo{author}{\bibfnamefont{G.}~\bibnamefont{Schehr}},
  \bibinfo{journal}{Adv. Phys.} \textbf{\bibinfo{volume}{62}},
  \bibinfo{pages}{225} (\bibinfo{year}{2013}).

\bibitem[{\citenamefont{Derrida}(1995)}]{Derrida95}
\bibinfo{author}{\bibfnamefont{B.}~\bibnamefont{Derrida}}, \bibinfo{journal}{J.
  Phys. A: Math. Gen.} \textbf{\bibinfo{volume}{28}}, \bibinfo{pages}{1481}
  (\bibinfo{year}{1995}).

\bibitem[{\citenamefont{Derrida et~al.}(1995)\citenamefont{Derrida, Hakim, and
  Pasquier}}]{DeHaPa95}
\bibinfo{author}{\bibfnamefont{B.}~\bibnamefont{Derrida}},
  \bibinfo{author}{\bibfnamefont{V.}~\bibnamefont{Hakim}}, \bibnamefont{and}
  \bibinfo{author}{\bibfnamefont{V.}~\bibnamefont{Pasquier}},
  \bibinfo{journal}{Phys. Rev. Lett.} \textbf{\bibinfo{volume}{75}},
  \bibinfo{pages}{751} (\bibinfo{year}{1995}).

\bibitem[{\citenamefont{Hinrichsen}(2000)}]{Hinrichsen00}
\bibinfo{author}{\bibfnamefont{H.}~\bibnamefont{Hinrichsen}},
  \bibinfo{journal}{Adv. Phys.}  (\bibinfo{year}{2000}).

\bibitem[{\citenamefont{Kang et~al.}(1991)\citenamefont{Kang, Weinberg, and
  Deem}}]{KaWeDe91}
\bibinfo{author}{\bibfnamefont{H.~C.} \bibnamefont{Kang}},
  \bibinfo{author}{\bibfnamefont{W.~H.} \bibnamefont{Weinberg}},
  \bibnamefont{and} \bibinfo{author}{\bibfnamefont{M.~W.} \bibnamefont{Deem}},
  \bibinfo{journal}{Phys. Rev. B} \textbf{\bibinfo{volume}{43}},
  \bibinfo{pages}{11438} (\bibinfo{year}{1991}).

\bibitem[{\citenamefont{Karma and Lobkovsky}(2005)}]{KaLo05}
\bibinfo{author}{\bibfnamefont{A.}~\bibnamefont{Karma}} \bibnamefont{and}
  \bibinfo{author}{\bibfnamefont{A.~E.} \bibnamefont{Lobkovsky}},
  \bibinfo{journal}{Phys. Rev. E} \textbf{\bibinfo{volume}{71}},
  \bibinfo{pages}{036114} (\bibinfo{year}{2005}).

\bibitem[{\citenamefont{Cerf and Louhichi}(2007)}]{CeLo07}
\bibinfo{author}{\bibfnamefont{R.}~\bibnamefont{Cerf}} \bibnamefont{and}
  \bibinfo{author}{\bibfnamefont{S.}~\bibnamefont{Louhichi}},
  \bibinfo{journal}{Probab. Theory Relat. Fields}
  \textbf{\bibinfo{volume}{137}}, \bibinfo{pages}{379} (\bibinfo{year}{2007}).

\bibitem[{\citenamefont{Arenzon et~al.}(2015)\citenamefont{Arenzon,
  Cugliandolo, and Picco}}]{ArCuPi15}
\bibinfo{author}{\bibfnamefont{J.~J.} \bibnamefont{Arenzon}},
  \bibinfo{author}{\bibfnamefont{L.~F.} \bibnamefont{Cugliandolo}},
  \bibnamefont{and} \bibinfo{author}{\bibfnamefont{M.}~\bibnamefont{Picco}},
  \bibinfo{journal}{Phys. Rev. E} \textbf{\bibinfo{volume}{91}},
  \bibinfo{pages}{032142} (\bibinfo{year}{2015}).

\bibitem[{\citenamefont{Derrida et~al.}(1994)\citenamefont{Derrida, Bray, and
  Godrèche}}]{DeBrGo94}
\bibinfo{author}{\bibfnamefont{B.}~\bibnamefont{Derrida}},
  \bibinfo{author}{\bibfnamefont{A.~J.} \bibnamefont{Bray}}, \bibnamefont{and}
  \bibinfo{author}{\bibfnamefont{C.}~\bibnamefont{Godrèche}},
  \bibinfo{journal}{J. Phys. A: Math. Gen.} \textbf{\bibinfo{volume}{27}},
  \bibinfo{pages}{L357} (\bibinfo{year}{1994}).

\bibitem[{\citenamefont{Stauffer}(1994)}]{Stauffer94}
\bibinfo{author}{\bibfnamefont{D.}~\bibnamefont{Stauffer}},
  \bibinfo{journal}{J. Phys. A: Math. Gen.} \textbf{\bibinfo{volume}{27}},
  \bibinfo{pages}{5029} (\bibinfo{year}{1994}).

\bibitem[{\citenamefont{Blanchard
  et~al.}(2014{\natexlab{a}})\citenamefont{Blanchard, Cugliandolo, and
  Picco}}]{BlCuPi14}
\bibinfo{author}{\bibfnamefont{T.}~\bibnamefont{Blanchard}},
  \bibinfo{author}{\bibfnamefont{L.~F.} \bibnamefont{Cugliandolo}},
  \bibnamefont{and} \bibinfo{author}{\bibfnamefont{M.}~\bibnamefont{Picco}},
  \bibinfo{journal}{J. Stat. Mech.} p. \bibinfo{pages}{P12021}
  (\bibinfo{year}{2014}{\natexlab{a}}).

\bibitem[{\citenamefont{Howard and Godrèche}(1998)}]{HoGo98}
\bibinfo{author}{\bibfnamefont{M.}~\bibnamefont{Howard}} \bibnamefont{and}
  \bibinfo{author}{\bibfnamefont{C.}~\bibnamefont{Godrèche}},
  \bibinfo{journal}{J. Phys. A: Math. Gen.} \textbf{\bibinfo{volume}{31}},
  \bibinfo{pages}{L209} (\bibinfo{year}{1998}).

\bibitem[{\citenamefont{Derrida et~al.}(1996)\citenamefont{Derrida,
  de~Oliveira, and Stauffer}}]{DeOlSt96}
\bibinfo{author}{\bibfnamefont{B.}~\bibnamefont{Derrida}},
  \bibinfo{author}{\bibfnamefont{P.~M.~C.} \bibnamefont{de~Oliveira}},
  \bibnamefont{and} \bibinfo{author}{\bibfnamefont{D.}~\bibnamefont{Stauffer}},
  \bibinfo{journal}{Physica A} \textbf{\bibinfo{volume}{224}},
  \bibinfo{pages}{604} (\bibinfo{year}{1996}).

\bibitem[{\citenamefont{Lipowski}(1999)}]{Lipowski99}
\bibinfo{author}{\bibfnamefont{A.}~\bibnamefont{Lipowski}},
  \bibinfo{journal}{Physica A} \textbf{\bibinfo{volume}{268}},
  \bibinfo{pages}{6} (\bibinfo{year}{1999}).

\bibitem[{\citenamefont{Spirin et~al.}(2001{\natexlab{a}})\citenamefont{Spirin,
  Krapivsky, and Redner}}]{SpKrRe01a}
\bibinfo{author}{\bibfnamefont{V.}~\bibnamefont{Spirin}},
  \bibinfo{author}{\bibfnamefont{P.~L.} \bibnamefont{Krapivsky}},
  \bibnamefont{and} \bibinfo{author}{\bibfnamefont{S.}~\bibnamefont{Redner}},
  \bibinfo{journal}{Phys. Rev. E} \textbf{\bibinfo{volume}{63}},
  \bibinfo{pages}{036118} (\bibinfo{year}{2001}{\natexlab{a}}).

\bibitem[{\citenamefont{Spirin et~al.}(2001{\natexlab{b}})\citenamefont{Spirin,
  Krapivsky, and Redner}}]{SpKrRe01b}
\bibinfo{author}{\bibfnamefont{V.}~\bibnamefont{Spirin}},
  \bibinfo{author}{\bibfnamefont{P.~L.} \bibnamefont{Krapivsky}},
  \bibnamefont{and} \bibinfo{author}{\bibfnamefont{S.}~\bibnamefont{Redner}},
  \bibinfo{journal}{Phys. Rev. E} \textbf{\bibinfo{volume}{65}},
  \bibinfo{pages}{016119} (\bibinfo{year}{2001}{\natexlab{b}}).

\bibitem[{\citenamefont{Barros et~al.}(2009)\citenamefont{Barros, Krapivsky,
  and Redner}}]{BaKrRe09}
\bibinfo{author}{\bibfnamefont{K.}~\bibnamefont{Barros}},
  \bibinfo{author}{\bibfnamefont{P.~L.} \bibnamefont{Krapivsky}},
  \bibnamefont{and} \bibinfo{author}{\bibfnamefont{S.}~\bibnamefont{Redner}},
  \bibinfo{journal}{Phys. Rev. E} \textbf{\bibinfo{volume}{80}},
  \bibinfo{pages}{040101(R)} (\bibinfo{year}{2009}).

\bibitem[{\citenamefont{Olejarz et~al.}(2012)\citenamefont{Olejarz, Krapivsky,
  and Redner}}]{OlKrRe12}
\bibinfo{author}{\bibfnamefont{J.}~\bibnamefont{Olejarz}},
  \bibinfo{author}{\bibfnamefont{P.~L.} \bibnamefont{Krapivsky}},
  \bibnamefont{and} \bibinfo{author}{\bibfnamefont{S.}~\bibnamefont{Redner}},
  \bibinfo{journal}{Phys. Rev. Lett.} \textbf{\bibinfo{volume}{109}},
  \bibinfo{pages}{195702} (\bibinfo{year}{2012}).

\bibitem[{\citenamefont{Godrèche and Pleimling}(2018)}]{GoPl18}
\bibinfo{author}{\bibfnamefont{C.}~\bibnamefont{Godrèche}} \bibnamefont{and}
  \bibinfo{author}{\bibfnamefont{M.}~\bibnamefont{Pleimling}},
  \bibinfo{journal}{J. Stat. Mech.} p. \bibinfo{pages}{043209}
  (\bibinfo{year}{2018}).

\bibitem[{\citenamefont{Arenzon et~al.}(2007)\citenamefont{Arenzon, Bray,
  Cugliandolo, and Sicilia}}]{ArBrCuSi07}
\bibinfo{author}{\bibfnamefont{J.~J.} \bibnamefont{Arenzon}},
  \bibinfo{author}{\bibfnamefont{A.~J.} \bibnamefont{Bray}},
  \bibinfo{author}{\bibfnamefont{L.~F.} \bibnamefont{Cugliandolo}},
  \bibnamefont{and} \bibinfo{author}{\bibfnamefont{A.}~\bibnamefont{Sicilia}},
  \bibinfo{journal}{Phys. Rev. Lett.} \textbf{\bibinfo{volume}{98}},
  \bibinfo{pages}{145701} (\bibinfo{year}{2007}).

\bibitem[{\citenamefont{Sicilia et~al.}(2007)\citenamefont{Sicilia, Arenzon,
  Bray, and Cugliandolo}}]{SiArBrCu07}
\bibinfo{author}{\bibfnamefont{A.}~\bibnamefont{Sicilia}},
  \bibinfo{author}{\bibfnamefont{J.~J.} \bibnamefont{Arenzon}},
  \bibinfo{author}{\bibfnamefont{A.~J.} \bibnamefont{Bray}}, \bibnamefont{and}
  \bibinfo{author}{\bibfnamefont{L.~F.} \bibnamefont{Cugliandolo}},
  \bibinfo{journal}{Phys. Rev. E} \textbf{\bibinfo{volume}{76}},
  \bibinfo{pages}{61116} (\bibinfo{year}{2007}).

\bibitem[{\citenamefont{Blanchard
  et~al.}(2014{\natexlab{b}})\citenamefont{Blanchard, Corberi, Cugliandolo, and
  Picco}}]{BlCoCuPi14}
\bibinfo{author}{\bibfnamefont{T.}~\bibnamefont{Blanchard}},
  \bibinfo{author}{\bibfnamefont{F.}~\bibnamefont{Corberi}},
  \bibinfo{author}{\bibfnamefont{L.~F.} \bibnamefont{Cugliandolo}},
  \bibnamefont{and} \bibinfo{author}{\bibfnamefont{M.}~\bibnamefont{Picco}},
  \bibinfo{journal}{Europhys. Lett.} \textbf{\bibinfo{volume}{106}},
  \bibinfo{pages}{66001} (\bibinfo{year}{2014}{\natexlab{b}}).

\bibitem[{\citenamefont{Blanchard et~al.}(2017)\citenamefont{Blanchard,
  Cugliandolo, Picco, and Tartaglia}}]{BlCuPiTa17}
\bibinfo{author}{\bibfnamefont{T.}~\bibnamefont{Blanchard}},
  \bibinfo{author}{\bibfnamefont{L.~F.} \bibnamefont{Cugliandolo}},
  \bibinfo{author}{\bibfnamefont{M.}~\bibnamefont{Picco}}, \bibnamefont{and}
  \bibinfo{author}{\bibfnamefont{A.}~\bibnamefont{Tartaglia}},
  \bibinfo{journal}{J. Stat. Mech.} p. \bibinfo{pages}{113201}
  (\bibinfo{year}{2017}).

\bibitem[{\citenamefont{de~Azevedo-Lopes
  et~al.}(2022)\citenamefont{de~Azevedo-Lopes, Almeida, de~Oliveira, and
  Arenzon}}]{AzAlOlAr22}
\bibinfo{author}{\bibfnamefont{A.}~\bibnamefont{de~Azevedo-Lopes}},
  \bibinfo{author}{\bibfnamefont{R.~A.~L.} \bibnamefont{Almeida}},
  \bibinfo{author}{\bibfnamefont{P.~M.~C.} \bibnamefont{de~Oliveira}},
  \bibnamefont{and} \bibinfo{author}{\bibfnamefont{J.~J.}
  \bibnamefont{Arenzon}} (\bibinfo{year}{2022}), \bibinfo{note}{to be
  published}.

\bibitem[{\citenamefont{Grant and Gunton}(1983)}]{GrGu83}
\bibinfo{author}{\bibfnamefont{M.}~\bibnamefont{Grant}} \bibnamefont{and}
  \bibinfo{author}{\bibfnamefont{J.~D.} \bibnamefont{Gunton}},
  \bibinfo{journal}{Phys. Rev. B} \textbf{\bibinfo{volume}{28}},
  \bibinfo{pages}{5496} (\bibinfo{year}{1983}).

\bibitem[{\citenamefont{Sahni et~al.}(1983)\citenamefont{Sahni, Grest, and
  Safran}}]{SaGrSa83}
\bibinfo{author}{\bibfnamefont{P.~S.} \bibnamefont{Sahni}},
  \bibinfo{author}{\bibfnamefont{G.~S.} \bibnamefont{Grest}}, \bibnamefont{and}
  \bibinfo{author}{\bibfnamefont{S.~A.} \bibnamefont{Safran}},
  \bibinfo{journal}{Phys. Rev. Lett.} \textbf{\bibinfo{volume}{50}},
  \bibinfo{pages}{60} (\bibinfo{year}{1983}).

\bibitem[{\citenamefont{Masuda et~al.}(2010)\citenamefont{Masuda, Gibert, and
  Redner}}]{MaGiRe10}
\bibinfo{author}{\bibfnamefont{N.}~\bibnamefont{Masuda}},
  \bibinfo{author}{\bibfnamefont{N.}~\bibnamefont{Gibert}}, \bibnamefont{and}
  \bibinfo{author}{\bibfnamefont{S.}~\bibnamefont{Redner}},
  \bibinfo{journal}{Phys. Rev. E} \textbf{\bibinfo{volume}{82}},
  \bibinfo{pages}{010103(R)} (\bibinfo{year}{2010}).

\bibitem[{\citenamefont{Caccioli et~al.}(2013)\citenamefont{Caccioli,
  Dall’Asta, Galla, and Rogers}}]{CaDaGaRo13}
\bibinfo{author}{\bibfnamefont{F.}~\bibnamefont{Caccioli}},
  \bibinfo{author}{\bibfnamefont{L.}~\bibnamefont{Dall’Asta}},
  \bibinfo{author}{\bibfnamefont{T.}~\bibnamefont{Galla}}, \bibnamefont{and}
  \bibinfo{author}{\bibfnamefont{T.}~\bibnamefont{Rogers}},
  \bibinfo{journal}{Phys. Rev. E} \textbf{\bibinfo{volume}{87}},
  \bibinfo{pages}{052114} (\bibinfo{year}{2013}).

\bibitem[{\citenamefont{Sicilia et~al.}(2009)\citenamefont{Sicilia, Sarrazin,
  Arenzon, Bray, and Cugliandolo}}]{SiSaArNrCu09}
\bibinfo{author}{\bibfnamefont{A.}~\bibnamefont{Sicilia}},
  \bibinfo{author}{\bibfnamefont{Y.}~\bibnamefont{Sarrazin}},
  \bibinfo{author}{\bibfnamefont{J.~J.} \bibnamefont{Arenzon}},
  \bibinfo{author}{\bibfnamefont{A.~J.} \bibnamefont{Bray}}, \bibnamefont{and}
  \bibinfo{author}{\bibfnamefont{L.~F.} \bibnamefont{Cugliandolo}},
  \bibinfo{journal}{Phys. Rev. E} \textbf{\bibinfo{volume}{80}},
  \bibinfo{pages}{031121} (\bibinfo{year}{2009}).

\end{thebibliography}

\end{document}